\documentclass[fleqn,usenatbib]{mnras}

\usepackage{newtxtext,newtxmath}
\usepackage[T1]{fontenc}

\DeclareRobustCommand{\VAN}[3]{#2}
\let\VANthebibliography\thebibliography
\def\thebibliography{\DeclareRobustCommand{\VAN}[3]{##3}\VANthebibliography}

\usepackage{graphicx}	% Including figure files
\usepackage{amsmath}	% Advanced maths commands

\usepackage{pdflscape}	% Landscape pages
\usepackage{xcolor}
\title[Jet–Interstellar Medium Interactions in Gaseous Disks]{Jet–ISM Interactions in Gaseous Disks: Simulating Kinetic Feedback in the Radio Galaxy 3C\,326\,N}

\author[Shende et al.]{Mayur B. Shende,$^{1,2,3}$\thanks{E-mail: mayurshende@zju.edu.cn (MS)}
Dipanjan Mukherjee,$^{3}$
N. P. H. Nesvadba,$^{4}$
Geoffrey Bicknell,$^{5}$
James Leftley,$^{6}$
\newauthor
Moun Meenakshi,$^{7}$ 
and Raghav Gogia $^{3}$\\
$^{1}$Center for Cosmology and Computational Astrophysics, Institute for Advanced Study in Physics, Zhejiang University, Hangzhou 310058, China \\
$^{2}$Institute of Astronomy, School of Physics, Zhejiang University, Hangzhou 310058, China \\
$^{3}$Inter-University Centre for Astronomy and Astrophysics, Ganeshkhind, Savitribai Phule Pune University Campus, Pune- 411007, India\\
$^{4}$Universit\'e de la C\^ote d'Azur, Observatoire de la C\^ote d'Azur, CNRS, Laboratoire Lagrange, Bd de l'Observatoire, CS 34229, F-06304 Nice cedex 4, France\\
$^{5}$Research School of Astronomy and Astrophysics, The Australian National University, Canberra, ACT 2611, Australia\\
$^{6}$European Southern Observatory, Alonso de C\'ordova 3107, Vitacura Casilla 19001, Santiago, Chile\\
$^7$ Leibniz Institute for Astrophysics, An der Sternwarte 16, D-14482 Potsdam, Germany
}

\date{Accepted XXX. Received YYY; in original form ZZZ}

\pubyear{\the\year{}}

\begin{document}
\label{firstpage}
\pagerange{\pageref{firstpage}--\pageref{lastpage}}
\maketitle

\begin{abstract}
%We present 3D relativistic hydrodynamic simulations of jet--ISM interaction in an inhomogeneous gaseous disk, spanning a parameter space of cloud configurations, jet powers, and disk central densities. Our simulations incorporate a numerical scheme that prevents the unphysical disk collapse that limited earlier studies and maintains turbulence in the disk over time. We find that the outflow morphology, velocity dispersion, and kinetic energy depend strongly on the underlying cloud configuration. Simulations with small ISM clouds ($l_{\rm cmax}=50$~pc) exhibit the highest velocity dispersion and kinetic energy, in contrast to the large-cloud case ($l_{\rm cmax}=250$~pc), which shows the lowest values; the mixed configuration lies intermediately between the two. Consistent with previous studies, we find that outflow strength increases with increasing jet power and decreasing disk central density. Importantly, synthetic emission maps from our fiducial simulation (a $10^{45}\,\rm erg\,s^{-1}$ jet interacting with a mixed cloud configuration) successfully reproduce the observed jet-driven bubble morphology in the radio galaxy 3C~326~N, including its spatial structure and line-of-sight velocity profiles.

Several radio galaxies, such as 3C\,326\,N, show signatures of jet–ISM coupling, but a complete theoretical framework for explaining them is still lacking. Interpreting these observations requires a detailed understanding of the gas distribution, geometry, and outflow energetics. In this paper, we use three-dimensional relativistic hydrodynamic simulations to investigate jet--ISM coupling in inhomogeneous gaseous disks, exploring a parameter space spanning different cloud configurations, jet powers, and central disk densities. Our simulations incorporate a numerical turbulence injection scheme that maintains vertical support in the disk, preventing the unphysical collapse encountered in previous studies. We find that jet--ISM coupling is strongly governed by the underlying cloud configuration, leading to distinct outflow morphologies, velocity dispersions, and kinetic energies. Simulations with small-scale ($l_{\rm c,max}=50$~pc) clouds produce the highest velocity dispersions and kinetic energies, whereas large-scale cloud configurations ($l_{\rm c,max}=250$~pc) yield the lowest values, with mixed cloud distributions exhibiting intermediate behavior. In addition, mixed cloud configurations give rise to asymmetric jet propagation, naturally producing unequal lobe lengths similar to those observed in radio galaxies. We compare our fiducial simulation (a $10^{45}\,\rm erg\,s^{-1}$ jet interacting with a mixed cloud configuration) with observations of 3C\,326\,N, focusing on the morphology of the jet-driven bubble, synthetic emission and the gas kinematics. Our results successfully reproduce the observed properties, providing strong evidence that jet--ISM interactions can account for the wide bubble and the complex gas kinematics observed in this system.

\end{abstract}


%%%%%%%%%%%%%%%%%%%%%%%%%%%%%%%%%%%%%%%%%%%%%%%%%%

%%%%%%%%%%%%%%%%% BODY OF PAPER %%%%%%%%%%%%%%%%%%

\section{Introduction}
It has now been well established that energy output from an Active Galactic Nucleus (AGN) -- whether as radiation or powerful relativistic jets -- strongly impacts galaxy evolution \citep[e.g. see reviews by][]{netzer_review_2015, padovani_review_2017, mukherjee_2025_review}. Such an energy output can have a profound impact on the surrounding medium by driving large-scale outflows and influencing the thermal, kinematic, and chemical state of the gas on galactic scales, as well as heating extra-galactic gas reservoirs to quenching accretion. Such proposed feedback effects have been supported by wide-ranging observations over the past few decades \citep[see reviews by][and references therein]{fabian_2012,harrison24a}. Early studies of the \emph{radio/maintenance} mode of AGN feedback primarily considered relativistic jets to be important for heating the large-scale gas reservoir surrounding galaxies \citep{bourne_and_yang_2023}. 
However, recent studies have now established that such jets can also have a substantial impact on the host galaxy's Interstellar Medium (ISM), as confirmed both from theoretical simulations, as well as wide variety of observed results \citep[see the review][ for a detailed discussion]{mukherjee_2025_review}. 

A particularly striking class of sources where this form of feedback is manifested are the so-called ``${\rm H_{2}}$ luminous'' radio galaxies, or MOlecular Hydrogen Emission Galaxies (MOHEGs). Observations with Spitzer/IRS reveal that 30\% of nearby ($z < 0.2$) powerful radio galaxies in the 3CRR catalogue fall into this category, showing bright, infrared line emission from ${\rm H_{2}}$ molecules, which does not appear to be associated with star formation \citep{3c326_ogle_2007, 3c326_ogle_2010, 3c326_nesvadba_2010, 3c326_nesvadba_2011}. The spectra are dominated by pure rotational lines of molecular hydrogen ($L({\rm H}_{2}) = 10^{40}-10^{43} \,\,{\rm erg\,\,s^{-1}}$), while the common star-formation tracers like bright IR continuum, mid-IR lines of [NeII] and [NeIII], and Polycyclic aromatic hydrocarbon (PAH) lines are either very weak or absent. Furthermore, it is also observed that these sources have radiatively weak AGN \citep{lanz_2016}. As a result, the ${\rm H_{2}}$ luminosity exceeds alternative molecular gas heating mechanisms like star formation or AGN radiation in extreme cases. On these strong physical grounds, \cite{3c326_ogle_2007} and \cite{3c326_nesvadba_2010, 3c326_nesvadba_2011} suggested that the bright $\rm H_{2}$ line emission is produced by the shocks created by the interaction of radio jets with the multiphase ISM of the host galaxy.

In the last two decades there have been a series of papers \citep[e.g.][ etc.]{sutherland_2007,wagner_2011,gaibler12a,wagner_2012, mukherjee_2016, mukherjee_2018,tanner_and_weaver_2022} that have simulated the interaction of a jet with an inhomogeneous ISM on kpc-scales. Such works have demonstrated how the jets pass through an initial `flood and channel' phase, where they couple strongly with the ISM, and drive strong multi-phase outflows ($\sim 100-1000\,\,\rm km\,\,s^{-1}$). A comprehensive summary of such simulations and their results has been presented in the review \citet{mukherjee_2025_review}. Of the above, the two papers most relevant to this  work are: i) \citet{wagner_2012}, where relativistic jets were propagated through an initially static, spherical shaped ISM. The paper explored a wide parameter space with with clouds of different length scales, densities and jet powers. ii) \citet{mukherjeeIC5063_2018,mukherjee_2018}, where jets of different powers and inclination interacted with kpc scale gas disks with varying mean densities. Some of the key results from these studies are: (a) Relativistic jets couple strongly with the inner parts of the disk. (b) Jets that are inclined towards the disk have a strong impact on it, launching sub-relativistic, wide angled outflows. (c) Shocks driven by the jet increase the velocity dispersion of gas inside the disk. (e) There is a strong dependence on cloud size, with ISM having larger cloud sizes show lowest mean outflow speeds.

However, there were several important aspects which have not been addressed. Firstly, \citet{mukherjee_2018} 
 does not explore the impact of jet interaction with varying cloud configurations within the disk. In their simulations, the maximum cloud size is fixed at approximately 150 pc. While this scale is broadly representative of typical giant molecular clouds (GMCs) observed in the interstellar medium (10–100 pc) in the Milky Way, it does not account for larger structures such as giant molecular associations (GMAs), which are commonly found in gas-rich galaxies.
%which are commonly found in the central regions of barred spiral galaxies and in minor mergers. 
Observations indicate that GMAs in these environments typically span $\sim$200--300 pc; examples include average sizes of $\sim$220 pc in barred spirals such as NGC 1365 and NGC 1097 \citep{sakamoto_2007, hsieh_2011}, and $\sim$250--300 pc in minor mergers like NGC 1614 and NGC 4194 \citep{olsson_2010, konig_2013, aalto_2000}.  Hence, the need to explore multi-scale cloud configurations in new simulations arises naturally from these observations. Although \citet{wagner_2012} performed an extensive suite of simulations with varying cloud sizes, the maximum cloud sizes explored were limited to $\sim 50$ pc.     

In addition, the previous simulations do not take into account the decay of turbulence in the disk over time when no jet is injected. This is because other feedback processes, like stellar and supernova feedback, are not included in their setup -- though these could be important for providing vertical support to the disk. Numerical diffusion also contributes to the decay. As a result, the disk collapses in about 3--4 Myr due to the lack of persistent turbulent support. Even with jet injection, the simulations cannot be run indefinitely, as that would no longer reflect a physically realistic scenario. This implies the necessity of developing a method by which turbulence can be maintained in order to prevent the unphysical collapse of the disk. In a recent paper \citep{borodina_2025}, a turbulence forcing function based on the Ornstein-Uhlenbeck (OU) process \citep{federrath_2008, federrath_2009} has been employed in simulations of jet-gas feedback. However, a drawback of such an implementation is the assumption of an unphysical periodic boundary condition at the outer boundaries, which limits the interpretation of the results from the simulation to the central kpc.

% There is a wide body of literature devoted to the study of ISM turbulence. \cite{mac_low_1998, stone_1998, ostriker_1999, padoan_and_nordlund_1999} study the turbulence decay rates in ISM clouds by performing HD and MHD simulations of supersonic, and super-Alfv\'{e}nic turbulence, and find that the kinetic energy decreases as a power-law in time. To sustain the turbulence in the ISM, \cite{federrath_2008, federrath_2009, federrath_2010, schmidt_2009} in their high-resolution simulations, solve fluid equations with a turbulence forcing term modeled by Ornstein-Uhlenbeck (OU) process. 

In this paper we follow the same simulation framework as \cite{mukherjee_2018}, but improve upon the previously mentioned drawbacks by developing a numerical scheme to stabilize the disk and maintain turbulence in the disk over time. The details of the simulation set up, initial conditions of the ISM and jet parameters are presented in Sec.~(\ref{sec.setup}), and the newly proposed turbulence injection prescription in Appendix~(\ref{appendix_A}). Using this framework, we have carried out a suite of simulations of jet-ISM interaction, with varying cloud configuration, jet power, and mean density within the disk, whose results are presented in Sec.~(\ref{sec.results}). This extends the parameter space of the simulations presented earlier in \cite{mukherjee_2018}. 

%An important goal of these simulations is to qualitatively explain the observed signatures of jet–ISM interaction in the radio galaxy 3C\,326\,N \citep{3c326_nesvadba_2010, 3c326_leftley_2024}. 
As of now, we have sufficiently rich and complete observational data to infer the physical processes at work in various sources. Observations can therefore help validate simulations across a wide parameter space. With this in mind, we wish to test whether we can successfully reproduce the observed morphology of jet-ISM interactions in the radio galaxy 3C\,326\,N \citep{3c326_nesvadba_2010, 3c326_leftley_2024}. In the recent past, several such joint comparison of observed results with simulations of jet-ISM interactions \citep[e.g.][]{mukherjeeIC5063_2018,zovaro19a,murthy_2022,audibert22a,fabbiano22a,murthy25a} have provided valuable insights on the nature of the interaction in such systems. 3C\,326\,N is an outstanding source, in that it presents a particularly clear example of jet-driven AGN feedback without major contamination from star formation or radiative AGN feedback. It also has an excellent, high spatial-resolution imaging spectroscopic data obtained with JWST NIRSpec and MIRI, including the ro-vibrational and pure-rotational lines of warm molecular Hydrogen. The data reach spatial resolutions of-order 100 pc at excellent signal-to-noise ratios, akin to the sizes of giant-molecular cloud complexes, which makes them particularly valuable for a detailed comparison with hydrodynamic simulations like ours. Motivated by such an approach, we compare the recent spatially resolved observations of 3C\,326\,N \citep{3c326_leftley_2024} with predictions from the simulations in Sec.~(\ref{sec.3C326}). Our conclusions are presented in Sec.~(\ref{sec.conclusions}).

\section{Simulation setup}\label{sec.setup}

\begin{figure*}
	\includegraphics[scale=0.49]{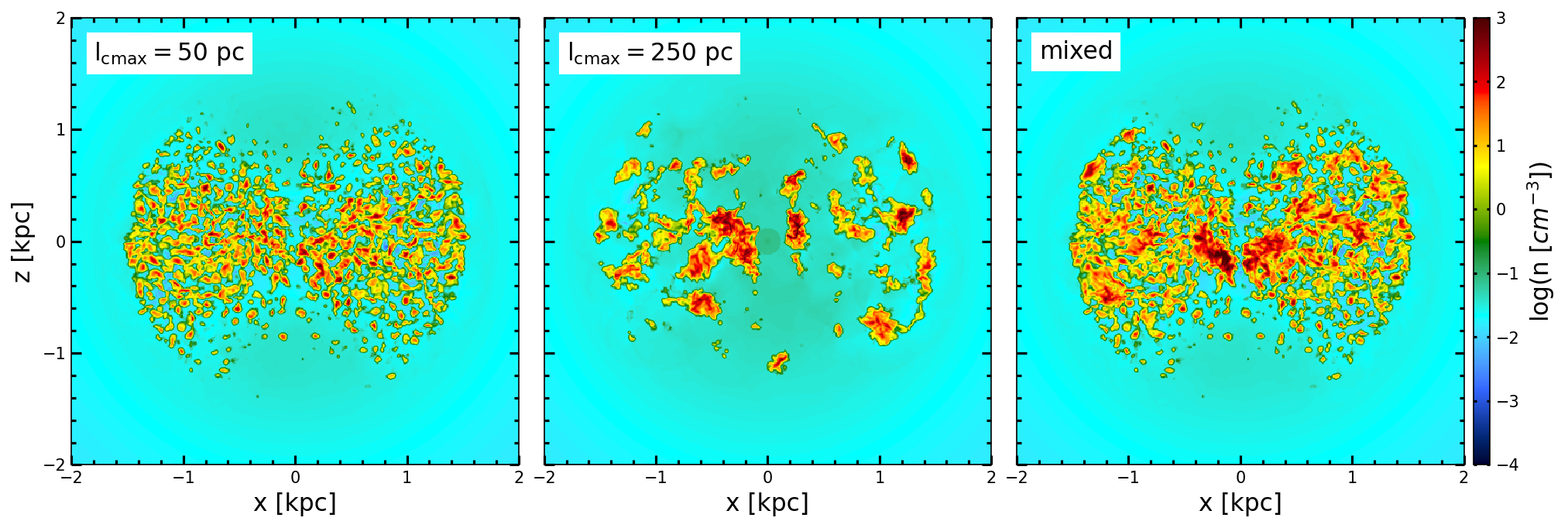}
    \caption{Density slices (log(n [$\rm cm^{-3}$])) in the $X-Z$ plane are shown for disks with different cloud configurations, taken prior to jet injection.
    }
    \label{fig:rho_lc_50_250_mixed_xz_ini}
\end{figure*}

We study the interaction between the relativistic jets and a turbulent gaseous disk using three-dimensional hydrodynamic simulations performed with the relativistic hydrodynamics module (RHD) of the publicly available \textsc{PLUTO} code \citep{mignone_2007}. The simulations were performed on a Cartesian domain with the physical dimensions of $\rm 4~ kpc\times4~kpc\times4~kpc$, on a grid of $512\times512\times512$ cells. The simulation setup is similar to that of \cite{mukherjee_2018}, with the inclusion of a newly developed numerical scheme to stabilize the disk and sustain the turbulence in the disk against decay (see Appendix~\ref{appendix_A}). A turbulent gaseous disk of radius $\sim 1.5\,\,\rm kpc$ is embedded in the hot isothermal halo ($T\sim 10^{7}\,\,\rm K$) in hydrostatic equilibrium. The halo number density profile is given by
\iffalse
\begin{equation}
    n_{\rm h} = n_{\rm h0}\,{\rm exp}\bigg\{\frac{-[\phi(r) - \phi(0)]}{k_{\rm B}T_{\rm h}/(\mu m_{\rm a})}\bigg\}    \label{eq:halo_density}
\end{equation}
\fi
\begin{equation}
    n_{\rm h} = n_{\rm h0}\,{\rm exp}\bigg\{- \frac{\mu m_{\rm a}}{k_{\rm B}T_{\rm h}}[\Phi(r) - \Phi(0)]\bigg\}    \label{eq:halo_density}
\end{equation}
where $n_{\rm h0}$ is the halo central density, $T_{\rm h}$ is the halo temperature ($\sim 10^{7}\,\,\rm K$), and $\Phi(r)$ is the external gravitational potential ($r$ being the spherical radius) following the double isothermal prescription of \cite{sutherland_2007} and \cite{mukherjee_2016}, which incorporates the contribution of baryonic (stellar) and dark components (see Table~\ref{tab:grav_parameters}). The mean molecular weight $\mu \sim 0.6$, the atomic mass unit $m_{\rm a} = 1.6605 \times 10^{-24}\,\, \rm g$, and $k_{\rm B}$ is the Boltzmann constant. 

The density of the warm disk is realized as a fractal with a single-point lognormal distribution, and a two-point power-law correlation with a Kolmogorov spectrum \citep{sutherland_2007, wagner_2011, wagner_2012}, which is created using the publicly available pyFC routine.\footnote{Available at \href{https://bitbucket.org/pandante/pyfc}{https://bitbucket.org/pandante/pyfc} (created and maintained by Dr. A. Y. Wagner)}
The lognormal probability density function of the density $\rho$ is given by 
\begin{equation}
    P(\rho) = \frac{1}{s \sqrt{2 \pi} \rho} \exp{\bigg[\frac{-({\rm ln\,\,\rho} - m)^2}{2 s^2}\bigg]}       \label{eq:lognormal}
\end{equation}
where $m = {\rm ln}(\mu^{2}/\sqrt{\sigma^{2} + \mu^{2}})$, and $s = \sqrt{{\rm ln}[(\sigma^{2}/\mu^{2}) + 1]}$, with $\mu$ and $\sigma^{2}$ being the mean and variance of the lognormal distribution.

In all previous applications of seeding an inhomogeneous ISM with a fractal distribution in simulations \citep[e.g.][etc.]{sutherland_2007, wagner_2011, mukherjee_2016, mukherjee_2018, bieri_2017, tanner_and_weaver_2022, ward24a}, a single lognormal distribution has been used to describe the entire ISM. Although a turbulent gas cloud is expected to have a lognormal density distribution \citep[for example, see][and references therein]{vazquez94a, li03a, federrath_2010, kritsuk17a, mandal24a}, the properties of such a distribution, e.g. its width are intricately related to the properties of the turbulent medium (Mach number and plasma beta). However,  the physical processes sustaining the turbulence, such as local shear and rate of supernovae explosions may vary across the ISM and will thus depend on the local conditions. Moreover, additional factors such as minor mergers or density waves may create temporary agglomeration of larger scale density structures which may exceed the sizes of typical GMCs, as mentioned earlier \citep{sakamoto_2007, hsieh_2011, olsson_2010, konig_2013, aalto_2000}.

Since we plan to study the effect of different cloud configurations, we initialise three different kinds of ISM: (i) a fractal cube having the average maximum cloud size: $l_{\rm cmax} = 50$ pc, (ii) a fractal cube with $l_{\rm cmax} = 250$ pc and (iii) a mixed configuration with contribution from both types of fractal cubes. The third case with mixed clouds also possesses maximum density structures of 250 pc, but with a larger filling factor than that of $l_{\rm cmax} = 250$ pc , as the potentially vacant regions in between two clouds are filled with smaller sized density structures (see bottom panel of Fig.~\ref{fig:rho_hist_lc_50_250_mixed}). Such a mixed configuration provides a more realistic description of an ISM with larger scale density correlations, without artificially vacant zones in between them (Fig.~\ref{fig:rho_lc_50_250_mixed_xz_ini}).

All fractal cubes were created using a dimensionless mean $\mu = 1$ and variance $\sigma^{2} = 40$ (see Eq.~\ref{eq:lognormal}) describing the lognormal distribution. Although higher than the $\sigma^{2}$ values previously considered by \cite{sutherland_2007} and \cite{mukherjee_2018}, we have chosen $\sigma^{2} = 40$ in order to have a smaller filling factors and to generate bigger size clouds with larger inter-cloud spacing than in previous studies. Following \citet{sutherland_2007}, the cubes are  apodized with a mean density $n_{\rm w0}$ to create a kpc scale gas disk: 
\begin{equation}
    \frac{n_{\rm w}(R, z)}{n_{\rm w0}} = {\rm exp}\bigg\{- \frac{1}{\sigma_{\rm t}^{2}} [\Phi(R, z) -e_{\rm K}^{2} \Phi(R, 0) - (1 - e_{\rm K}^{2}) \Phi(0, 0)] \bigg\},     \label{eq:disk_density}
\end{equation}
where $\sigma_t$ defines the scale height of the disk and $\Phi(R,z)$ is the external gravitational potential, modelled as a double-isothermal solution with contributions from both stellar baryonic component and dark matter \citep{sutherland_2007,mukherjee_2016}. Here $R$ is the cylindrical radius, and $z$ is the height from the disk mid-plane. The parameter $e_{\rm K}$ is a measure of sub-Keplerian rotation which the disk-shaped ISM is subjected to according to the relation
\begin{equation}
    v_{\phi} = e_{\rm K} \bigg( R \frac{\partial \Phi}{\partial R}\bigg)^{1/2},     \label{eq:azimuthal_velocity}
\end{equation}
with $e_{\rm K} = 0.93$ assumed for this work. Furthermore, an initial turbulent velocity modelled as a Gaussian distribution with dispersion $\sim 40\,\,\rm km\,\,s^{-1}$ along each Cartesian direction is added to the gas. 

To counteract the turbulent decay in the disk, we add a turbulent velocity field at regular intervals of time by importing from an externally generated file. The details of the implementation of the scheme to stabilize the disk is presented in  Appendix~(\ref{appendix_A}). Jets are injected as a bi-conical outflow with a half opening angle of $10^{\circ}$. The jet axis is inclined at $45^{\circ}$ from the $Z$-axis and  lies is in the $X$-$Z$ plane (the disk is in the $X$-$Y$ plane). The set up of the gas disk and the jet injection is identical to that used in \citet{mukherjee_2018}.
The simulation parameters are listed in Table~(\ref{tab:sim_parameters}).

\begin{table}
    \centering
    \caption{Parameters of the gravitational potential common to all simulations}
    \begin{tabular}{|l l |l|}
    \hline
         Parameters &  Symbol & Value\\
         \hline
         DM core radius & $\rm r_{D}$ & 10 kpc\\
         Baryonic velocity dispersion & $\rm \sigma_{B}$ & 250 km/s\\
         Ratio of DM to baryonic core radius & $\rm \lambda = r_{D}/r_{B}$ & 10\\
         Ratio of DM to baryonic velocity dispersion          & $\rm \kappa = \sigma_{D}/\sigma_{B}$    & 2\\
         \hline
    \end{tabular}
    \label{tab:grav_parameters}
\end{table}

\begin{table*}
    \centering
    \caption{Simulation parameters}
    \begin{tabular}{|l l |l|}
    \hline
         Parameters & Description  & Value\\
         \hline
         {\bf Box parameters} & & \\
         Box size  & & 4 kpc $\times$ 4 kpc $\times$ 4 kpc\\
         Grid size & & $512 \times 512 \times 512$ \\
         Resolution & & 7.8125 pc \\
         \hline
         {\bf Hot halo parameters} & &\\
         $T_{\rm h}$ & Temperature of hot atmosphere & $10^{7}$ K\\
         $n_{\rm h0}$ & Halo central density & 0.1 ${\rm cm^{-3}}$\\
         \hline
         {\bf Clouds and disk} & & \\
         $l_{\rm cmax}$ & Average maximum cloud size of a fractal cube & (50, 250) pc \\
         $R_{\rm disk}$ & Disk radius & 1.5 kpc \\
         $T_{\rm c}$ & Critical temperature of warm phase & $3 \times 10^{4}$ K \\
         $n_{\rm w0}$ & Mean central density of clouds & (10, 40) $\rm cm^{-3}$ \\ 
         $\sigma_{\rm t}$ & Combined turbulent and thermal dispersion & 40 ${\rm km\, s^{-1}}$ \\
         $e_{\rm K}$ & Ratio of azimuthal to Keplerian speed & 0.93 \\
         \hline
         {\bf Jet parameters} & & \\
         $P_{\rm jet}$ & Jet power & ($10^{44}, 10^{45}, 10^{46}$) ${\rm erg\,\,s^{-1}}$ \\ 
         $\Gamma$ & Jet bulk Lorentz factor at nozzle & 5 \\
         $\chi$ & Ratio of rest mass energy to relativistic enthalpy & 50 \\
         $R_{\rm jet}$ & Radius of jet inlet & 40 pc \\
         $\theta_{\rm jet}$ & Half opening angle of the jet & $10^{\circ}$ \\
         $\theta_{\rm inc}$ & Inclination of jet axis wrt Z-axis & $45^{\circ}$ \\
         \hline
    \end{tabular}
    \label{tab:sim_parameters}
\end{table*}

\begin{table}
    \centering
    \caption{List of simulations}
    \begin{tabular}{|l|l|l|l|l|}
    \hline
         Simulation & $P_{\rm jet}$ & Cloud & $n_{\rm w0}$ & $M_{\rm disk}$\\
         label & [$\rm erg\,\,s^{-1}$] & configuration & [$\rm cm^{-3}$] & [$M_{\odot}$] \\
         \hline
         P44-mix-n40 & $10^{44}$ & Mixed & 40 & $2.4\times10^{9}$\\
         P45-lc50-n40 & $10^{45}$ & $l_{\rm cmax} = 50\,\, \rm pc$ & 40 & $1.27\times10^{9}$\\
         P45-lc250-n40 & $10^{45}$ & $l_{\rm cmax} = 250\,\, \rm pc$ & 40 & $1.31\times10^{9}$\\
         P45-mix-n40 & $10^{45}$ & Mixed & 40 & $2.4\times10^{9}$\\
         P45-mix-n10 & $10^{45}$ & Mixed & 10 & $4.33\times10^{8}$\\
         P46-mix-n40 & $10^{46}$ & Mixed & 40 & $2.4\times10^{9}$\\
         
         \hline
    \end{tabular}
    \label{tab:simulation_list}
\end{table}

\section{Results}\label{sec.results}

We have performed a series of simulations with different cloud configurations ($l_{\rm cmax}$), jet powers ($P_{\rm jet}$), and mean central densities ($n_{\rm w0}$) (see Table~\ref{tab:simulation_list}). In the following sections, we examine how these parameters shape the jet–ISM interaction. We first discuss the evolution of disk morphology, followed by an analysis of gas dynamics using velocity dispersion, kinetic energy and turbulence quantification. We then investigate jet propagation and confinement, and finally characterize the multi-phase outflows produced in these interactions.

\subsection{Effect of different cloud configurations}

\subsubsection{Morphology of interaction and nature of outflow}\label{sec.gasmorph}

\begin{figure*}
	\includegraphics[scale=0.50]{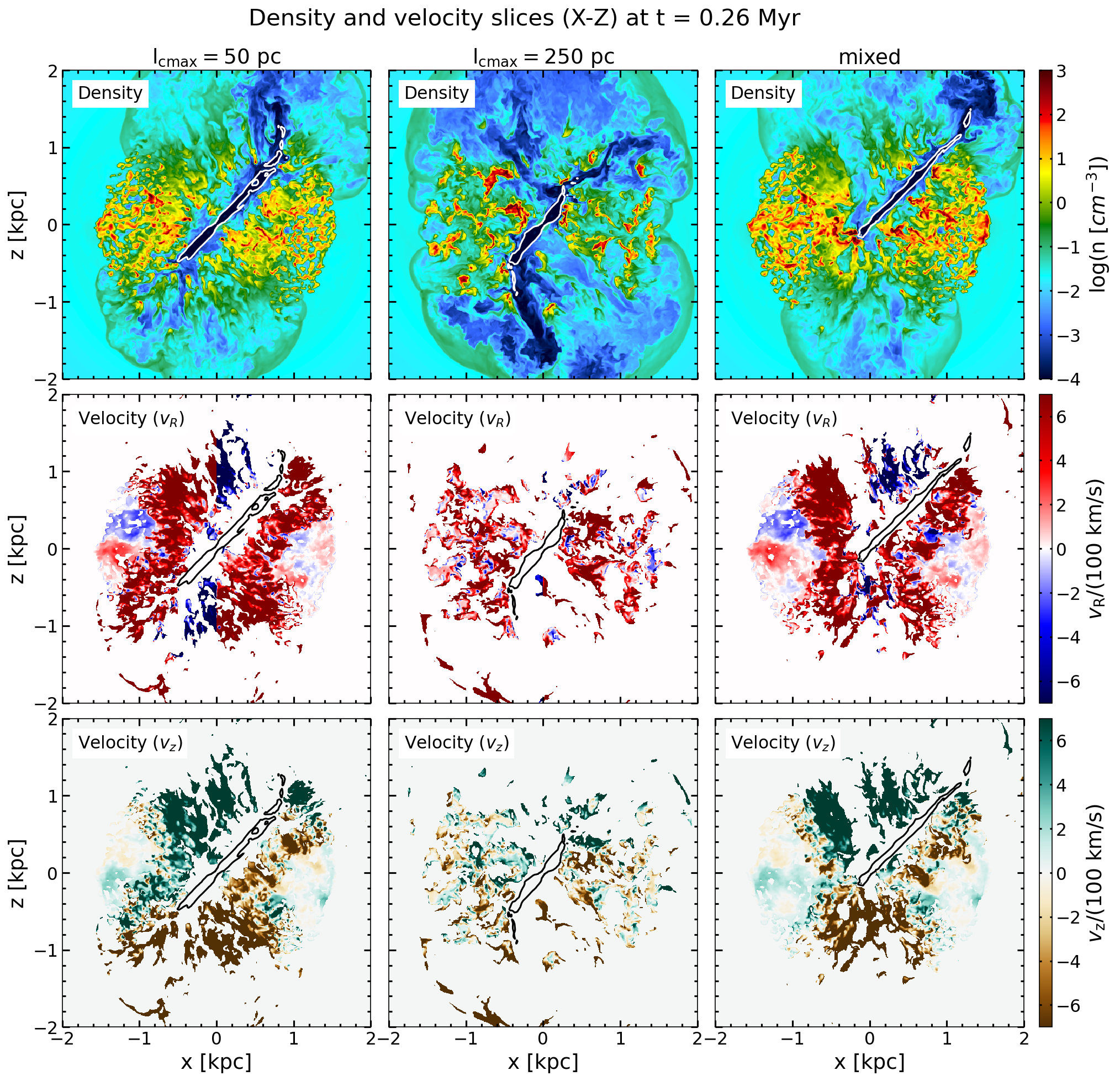}
    \caption{Density and velocity slices in the X-Z plane at 0.26 Myr for the simulations P45-lc50-n40 (\textit{left}), P45-lc250-n40 (\textit{middle}), and P45-mix-n40 (\textit{right}). Top: density ($\rm log\,\,n[cm^{-3}]$), middle and bottom: cylindrical radial velocity ($v_{\rm R}$), and vertical velocity ($v_{\rm z}$) normalized to 100 $\rm km\,\,s^{-1}$ for the dense gas (defined here as $n > 0.1\,\,{\rm cm^{-3}}$). The white and black contours in density and velocity profiles respectively, denote relativistic speeds with $\beta = 0.7$.}
    \label{fig:rho_vel_lc_50_250_mixed_xz_t1}
\end{figure*}

In this section we discuss the differences in nature of the interaction between a jet of power $P_{\rm jet} = 10^{45} \,\,\rm erg\,\,s^{-1}$ with an ISM of different cloud configurations (simulations P45-lc50-n40, P45-lc250-n40, and P45-mix-n40, respectively; see Table~\ref{tab:simulation_list}). 
% To investigate how jets interact with different cloud configurations, we simulate jets, with a power $P_{\rm jet} = 10^{45}\,\,\rm erg\,\,s^{-1}$, propagating into ISM disks characterized by cloud size distributions with $\lambda_{\rm max} = 100\,\,\rm pc$, $\lambda_{\rm max} = 500\,\,\rm pc$, and a configuration consisting of a mix of both scales (simulations P45-l100-n40, P45-l500-n40, and P45-mix-n40, respectively; see Table~\ref{tab:simulation_list} and Figure~\ref{fig:density_100_500_mixed_ini}). 
As discussed earlier in Sec.~(\ref{sec.setup}),  the $l_{\rm cmax} = 50\,\,\rm pc$ present smaller clouds which are more homogeneously distributed, as shown in Fig.~(\ref{fig:rho_lc_50_250_mixed_xz_ini}). The $l_{\rm cmax} = 250\,\,\rm pc$ have large clumps of gas, but also well separated with similar distances and hence larger inter-cloud distance. The mixed configuration on the other hand have smaller structures embedded within large scale clouds which fill up the inter-cloud voids. Hence, the three simulations probe three very different ISM configurations and the nature of jet-ISM interaction is also varied. Fig.~(\ref{fig:rho_vel_lc_50_250_mixed_xz_t1}) presents mid-plane density slices (top panel), and velocity profiles (middle and bottom panel) at 0.26 Myr, with each column denoting different cloud configurations. The white and black contours (in top panel and lower panels, respectively) indicate regions with velocity $\beta = 0.7$, highlighting the path of the jet.
%Figure~(\ref{fig:rho_vel_lc_50_250_mixed_xz_t1t2}) presents mid-plane density slices (top two panels) at 0.26 Myr (first row) and 0.81 Myr (second row), with each column denoting different cloud configurations. The white contours indicate regions with velocity $\beta = 0.7$, highlighting the path of the jet. 
\begin{itemize}
\item \emph{\underline{Smaller clouds ($l_{\rm cmax} = 50\,\, pc$):}}
We notice that the jet encounters less hindrance in simulations with an ISM disk characterized by $l_{\rm cmax} = 50\,\,\rm pc$, owing to the smaller cloud sizes. Since the smaller cloud sizes and consequently smaller cores are more easily disrupted, they lead to more widespread outflows near the path of the jet. This is shown in the middle-left and bottom-left panels of Fig.~(\ref{fig:rho_vel_lc_50_250_mixed_xz_t1}), which depicts the radial ($v_{\rm R}$) and vertical velocity ($v_{\rm z}$) profiles of dense gas (defined here as the gas with density $n > 0.1\,\,\rm cm^{-3}$). The amount of outflowing gas (which we define as the dense gas with $v_{\rm R} > 280\,\,\rm km \,\,s^{-1}$; $M_{v_R>280}$)\footnote{The threshold of $280\,\,\rm km\,\,s^{-1}$ is calculated from the $M_{\rm BH}-\sigma$ relation \citep[Eq. 1, and 19 of ][]{tremaine_2002} $$M_{\rm BH}/M_{\odot} = 10^{8.13}(\sigma_{200})^{4.02}$$ where $\sigma_{200}$ is stellar velocity dispersion in units of $200\,\,\rm km\,\,s^{-1}$. For a central black hole with $M_{\rm BH} = 5\times10^{8}M_{\odot}$ \citep[representative of 3C\,326\,N,][]{3c326_ogle_2007}, the dispersion turns out to be $\approx280\,\,\rm km\,\,s^{-1}$.} for this configuration at 0.26 Myr is found to be $\sim1.58\times10^{8}\,\, M_{\odot}$. The corresponding mass outflow rate is found to be  $\sim 388~M_{\odot}\,\,\rm yr^{-1}$.\footnote{Mass outflow rate here is defined as the ratio of the difference of outflow mass at two epochs to the time interval between those epochs.} As the jet pushes against the nearly homogeneous ISM, the outer edges are partially shielded by the formation of a dense wall along the jet. Hence, the outer edges show some inflowing structures.

\item \emph{\underline{Larger clouds  ($l_{\rm cmax} = 250\,\,pc$):} }
In contrast, in simulations with $l_{\rm cmax} = 250$ pc, the jet often gets stalled near larger clouds and takes a longer time to break out of the disk. However, the stalling of the jet-head leads to more flood-channel flows percolating in between the gaps of the larger density structures. This results in a more spherical bubble-shaped density contour (Fig.~\ref{fig:rho_vel_lc_50_250_mixed_xz_t1}, top centre panel). The efficient percolation through the clouds lead to  predominantly outflowing signatures in the radial velocity structures, and  show very little inflowing gas. The outflowing gas mass ($M_{v_R>280}$) and mass outflow rate for this configuration are $\sim 5.45\times10^{7}\,\,M_{\odot}$ and $\sim208\,\,M_{\odot}\,\,\rm yr^{-1}$.

\item \emph{\underline{Mixed configuration:}} 
In the mixed configuration, which offers a more realistic scenario, both the above signatures can be seen due to the existence of a multi-scale clumpy medium. While one jet is stalled, the other jet penetrates easily carving a narrow channel (Fig.~\ref{fig:rho_vel_lc_50_250_mixed_xz_t1}, right panel). They show strong wide-spread radial outflows both along the jet, as well as away from it, amounting to $M_{v_R>280} \approx 2.11\times10^{8}\,\,M_{\odot}$ with the outflow rate of $\sim 508~M_{\odot}\,\,\rm yr^{-1}$. Additionally, similar to the case of smaller clouds, prominent inflowing structures are visible at the outer edge. Thus, while the larger density structures can slow down the advance if directly along the jet's path, the more homogeneous small scale clouds are pushed aside, forming a dense wall of outflowing gas, which in turn may shield the outer parts of the disk.

\end{itemize}

\subsubsection{Velocity dispersion and feedback energetics} \label{sec:veldisp}

\begin{figure*}
	\includegraphics[scale=0.45]{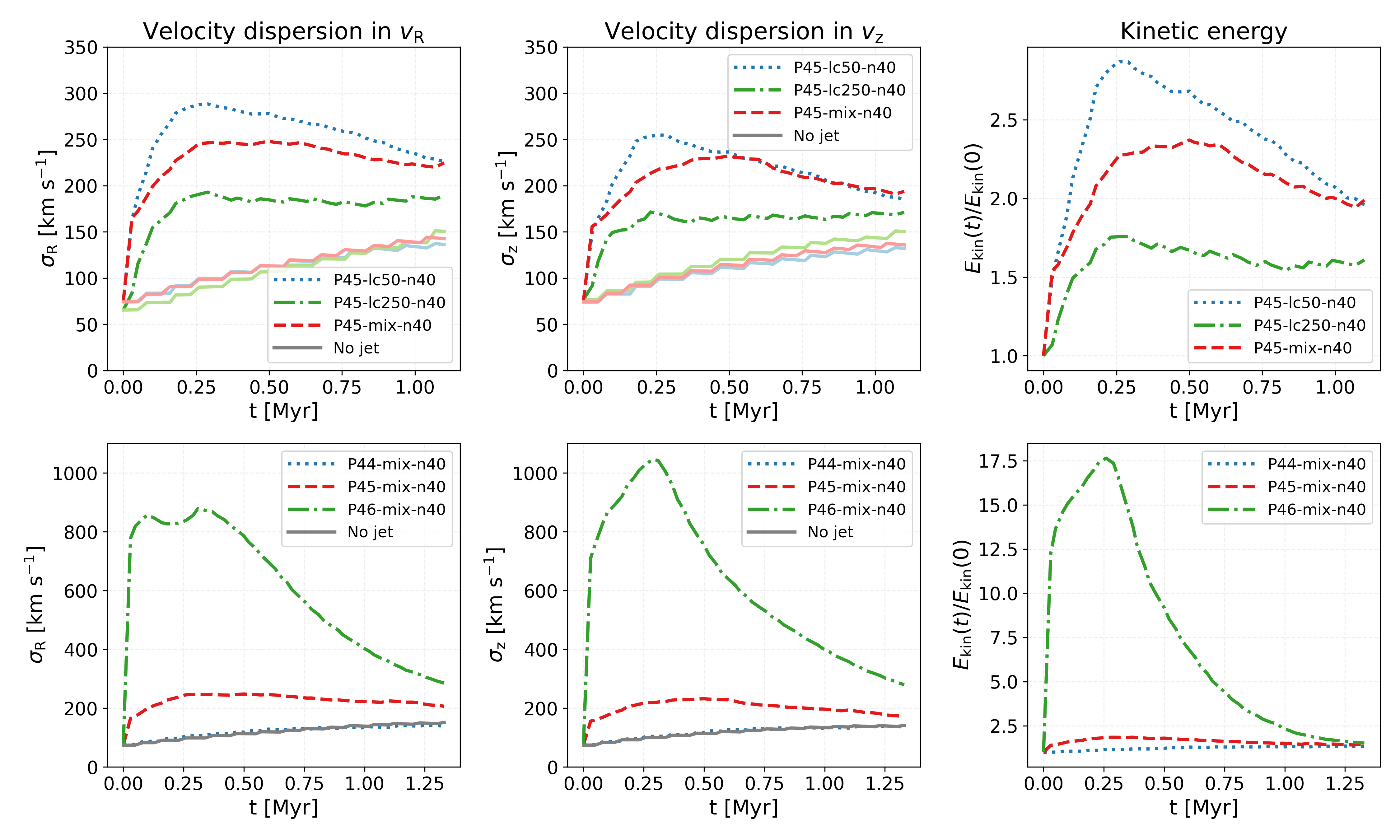}
    \caption{Top: Mass-weighted velocity dispersion ($\sigma_{\rm R}$: \textit{left}, $\sigma_{\rm z}$: \textit{middle}) and kinetic energy (\textit{right}) of dense gas (defined here as $n > 0.1\,\,\rm cm^{-3}$) for the simulations with different cloud configurations (see Sec.~\ref{sec:veldisp}). Bottom: Same as above but for simulations with different jet powers. Dependence on gas kinematics on jet power is discussed in \ref{sec:veldisp}. } 
    \label{fig:vel_dispersion_KE}
\end{figure*}

The trends discussed in the previous subsection are also supported by the evolution of the velocity dispersion of the dense gas. We calculate the $i$-th component of mass-weighted velocity dispersion ($\sigma_{i}$) using
\begin{equation}
    \sigma_{i}^{2} = \frac{1}{M} \Sigma \rho(v_{i} - \overline{v}_{i})^{2}, \,\,\,\,\,\,\,\,\,\,\,\,\,\,\,\,\,\,\,\,\,\overline{v}_{i} = \frac{1}{M} \Sigma\rho v_{i}   \label{eq:vel-dispersion}
\end{equation}
where the index $i$ corresponds to ($x, \,y, \,z$) components in Cartesian coordinates and ($R, \,\phi, \,z$) components in cylindrical coordinates. Here, $M$ is the total mass of the gas, $\rho$ is the gas density, $v_{i}$ is the $i$-th component of velocity, and $\overline{v}_{i}$ is its mass-weighted mean. 

The top panel of Fig.~(\ref{fig:vel_dispersion_KE}) shows the time evolution of mass-weighted velocity dispersion (radial and vertical components; $\sigma_{\rm R}$ and $\sigma_{\rm z}$, respectively), and kinetic energy of dense medium for all three cloud configurations, indicated by different line-styles, and with bright colours. The corresponding control runs with no jet are plotted with the lighter colours. As the jet begins to interact with the ISM, the velocity dispersion increases rapidly, reaching a peak that marks the phase of strongest jet-ISM interaction. It then declines when jet breaks out of the disk. We find that the peak of the velocity dispersion and the jet breakout time (defined later in Sec.~\ref{sec:jet_confinement}) are strongly correlated.

The simulation with smaller clouds ($l_{\rm cmax} = 50\,\,\rm pc$) shows the highest velocity dispersion, as the jet interacts more effectively with the low-mass clouds, dispersing them and stirring up the gas more efficiently. In contrast, ISM with larger clouds ($l_{\rm cmax} = 250\,\,\rm pc$) shows the lowest dispersion, since the denser and more massive cloud cores are more resistant to jet-induced disruption. The mixed configuration, containing clouds of both sizes, lies intermediate between the two cases, exhibiting moderate levels of velocity dispersion. Notably, in the $l_{\rm cmax} = 250\,\,\rm pc$ case, the dispersion in radial velocity plateaus at about 180 $\rm km \,\,s^{-1}$, indicating that the jet remains confined for a longer duration compared to the other two scenarios (see Sec.~\ref{sec:jet_confinement}).

\begin{figure}
	\includegraphics[scale=0.50]{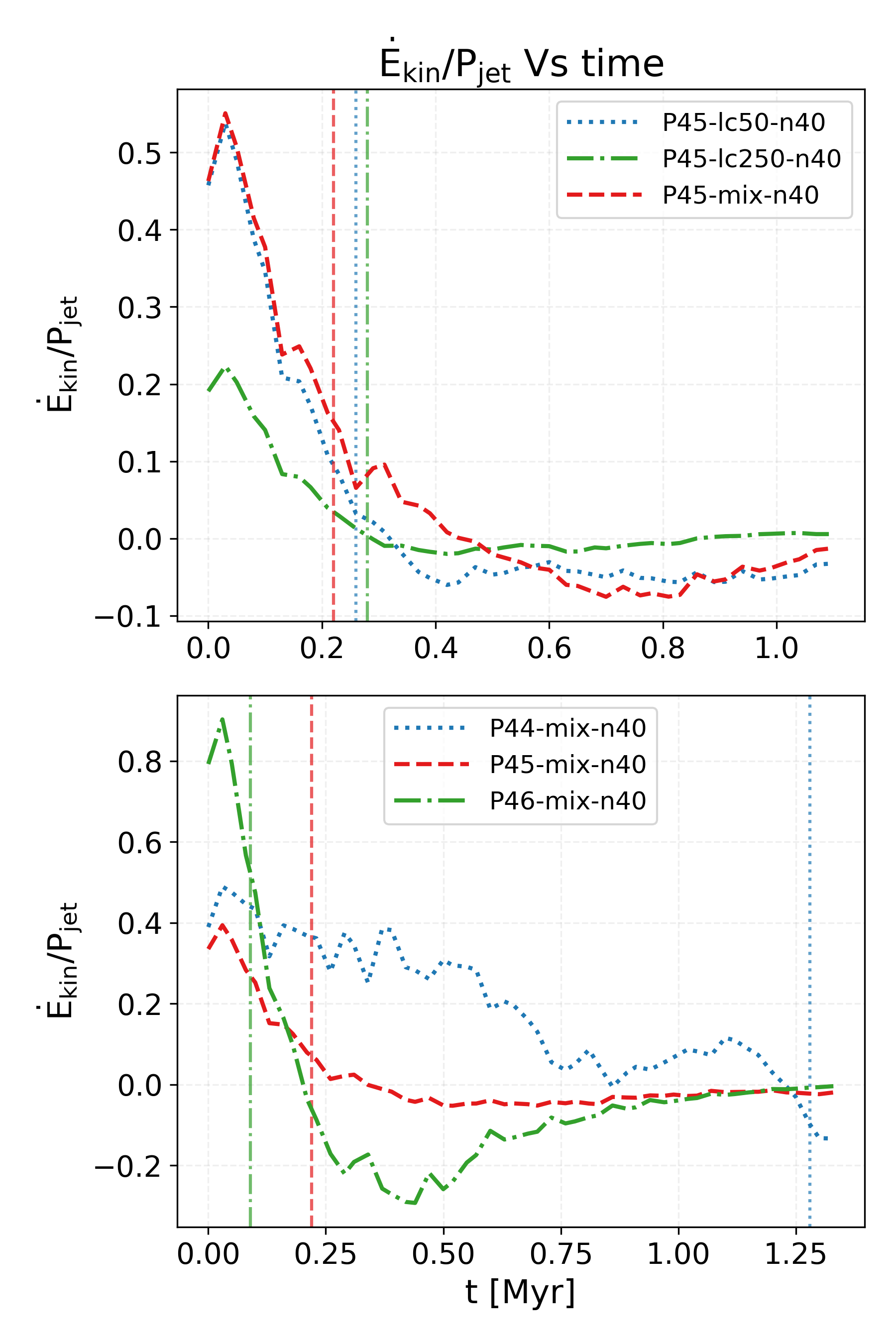}
    \caption{Rate of change of kinetic energy of dense gas as a function of time. The vertical lines correspond to breakout time of the forward jet (see Sec.~\ref{sec:veldisp}).}
    \label{fig:KE_jet_fraction}
\end{figure}

Similarly, as shown in the top right panel of Fig.~(\ref{fig:vel_dispersion_KE}), the kinetic energy imparted by the jet to the ISM disk is highest in the $l_{\rm cmax} = 50\,\,\rm pc$ configuration, followed by the mixed cloud case, and lowest in the $l_{\rm cmax} = 250\,\,\rm pc$ setup. 
%{\color{red} (Note to the reader: Following is the new analysis which I have added and is denoted in bold font.)} 
Another useful parameter to quantify the feedback energetics is to study the rate at which the jet imparts kinetic energy to the ISM with time. Fig.~(\ref{fig:KE_jet_fraction}) presents such an analysis where the ratio of the kinetic energy rate ($\dot{E}_{\rm kin}$) of dense ISM to the jet power ($P_{\rm jet}$) is plotted as a function of time. For reference, jet breakout times (defined later in Sec.~\ref{sec:jet_confinement}) for the forward jet are also shown (vertical lines). We notice that all the plots peak as soon as the jet starts to interact with the ISM, marking the initial transient phase of interaction, and is followed by the decrease in kinetic energy transfer rate. Ignoring the initial transient phase, we notice that the kinetic energy transfer rate around the jet-breakout time lie in the range $5-15$\% of the jet power for $l_{\rm cmax} = 50\,\,\rm pc$ and the mixed configuration, whereas for $l_{\rm cmax} = 250\,\,\rm pc$, it is $\lesssim5$\% (upper panel of Fig.~\ref{fig:KE_jet_fraction}). We emphasize here that the jet-breakout time is correlated with the maximum jet-ISM interaction. After some time, the kinetic energy rate of dense ISM becomes negative (at $\sim 0.3\,\,\rm Myr$ for $l_{\rm cmax} = 50,\, 250\,\,\rm pc$, and $\sim 0.45\,\,\rm Myr$ for mixed configuration), indicating that the ISM is losing kinetic energy. This is because of the significant decoupling of the jet from the ISM and eventual deceleration of the clouds, cooling and numerical diffusion.

\begin{figure*}
	\includegraphics[scale=0.55]{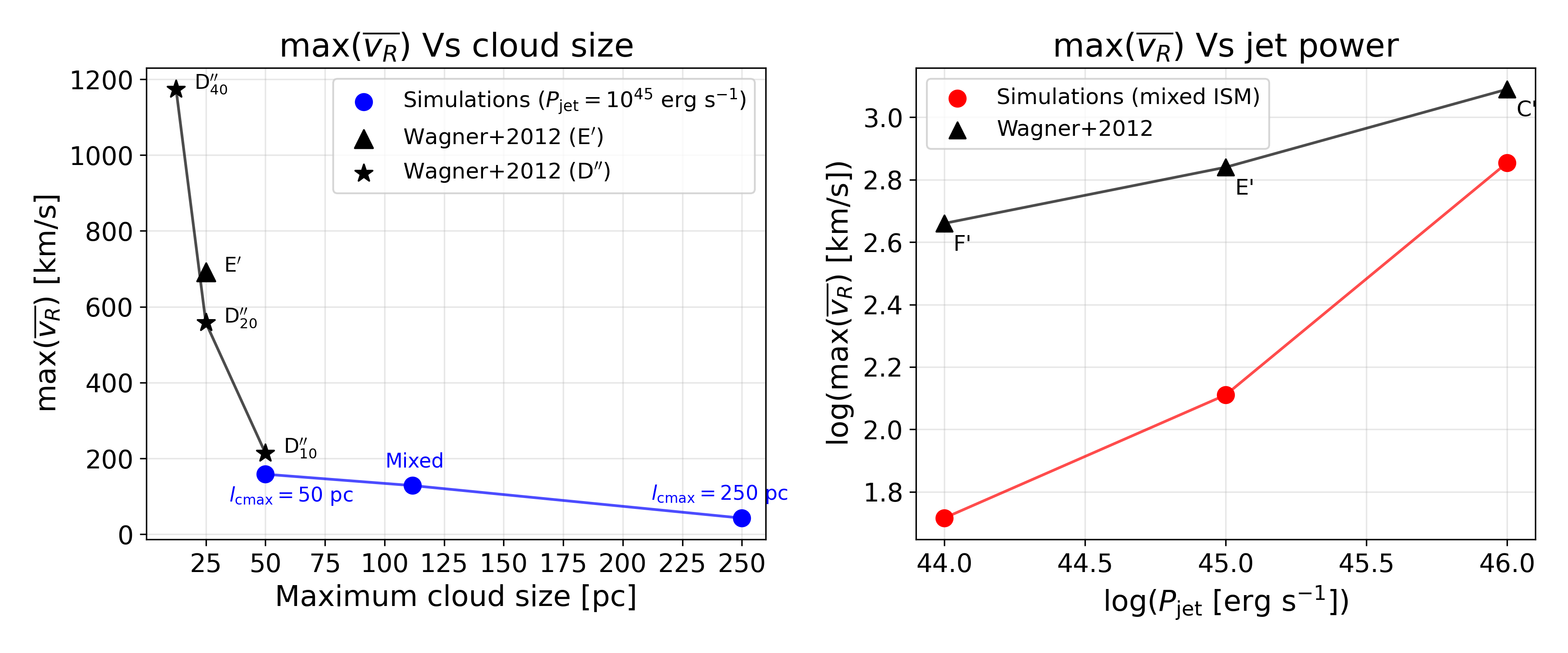}
    \caption{Left: Maximum mean radial velocity (mass-weighted) of clouds vs. maximum cloud size for the simulations with $P_{\rm jet} = 10^{45}\,\,\rm erg\,\,s^{-1}$. Simulations $\rm E^{\prime}$ and $\rm D^{\prime\prime}$ of \protect\cite{wagner_2012} are shown for comparison. Right: Maximum mean radial velocity (mass-weighted) against jet power. Results of \protect\cite{wagner_2012} are shown in filled triangles.}
    \label{fig:outflow_vR_lmax_power}
\end{figure*}

An important analysis performed by \cite{wagner_2011, wagner_2012} was to plot the maximum mass-weighted radial outflow velocities as a function of both the maximum cloud size in the ISM and the jet power. We have performed a similar analysis, shown in Fig.~(\ref{fig:outflow_vR_lmax_power}). For comparison with \citet[][hereafter WB12]{wagner_2012}, we select six simulations that have physical parameters close to ours. Simulations $\rm D^{\prime\prime}_{40}$, $\rm D^{\prime\prime}_{20}$ (which is listed as $\rm D^{\prime\prime}$ in WB12) and $\rm D^{\prime\prime}_{10}$ have same jet powers ($10^{45}\,\,\rm erg\,\,s^{-1}$), but different maximum cloud size configurations: 12.5, 25, and 50 pc, respectively. The mean density was $n_{\rm w0}=150\,\,\rm cm^{-3}$, which is higher than those in our simulations ($\sim 40\,\,\rm cm^{-3}$). The other three simulations, namely $\rm C^{\prime}$, $\rm E^{\prime}$, and $\rm F^{\prime}$, have the same maximum cloud size (25 pc) with mean density $n_{\rm w0}=30\,\,\rm cm^{-3}$, but different jet powers ($10^{44}$, $10^{45}$, and $10^{46}\,\,\rm erg\,\,s^{-1}$).

The left panel of Fig.~(\ref{fig:outflow_vR_lmax_power}) plots mass-weighted mean radial velocity $\overline{v_{\rm R}}$ as a function of the maximum cloud size. We consider the maximum value of $\overline{v_{\rm R}}$ among all the epochs of our simulations. We observe a nearly linear trend with our simulations, with $\overline{v_{\rm R}}$ decreasing as the cloud size increases.\footnote{For our mixed configuration, we compute an effective cloud size as the geometric mean of the two cloud sizes (50 pc and 250 pc): $(50\times250)^{1/2}\approx112$ pc.} The value of mean $\overline{v_{\rm R}}$ of our simulation with $l_{\rm cmax} = 50\,\,\rm pc$  approximately matches with $\rm D^{\prime\prime}_{10}$ of WB12 with a similar cloud size. However,  other simulations of WB12 with smaller clouds  ($\rm D^{\prime\prime}_{40}$, $\rm E^{\prime\prime}$, $\rm D^{\prime\prime}_{20}$) have much higher radial velocities.

There are several reasons for this discrepancy: 1) WB12 simulations cover the lower range of cloud sizes as compared to ours. As a result, they are prone to significantly higher values of acceleration and attaining high outflow velocities. 2) They use spherical geometry of cloud distribution which gives more coupling as compared to the disk in our case. 3) Their jet is four times more over-pressured relative to the ambient medium, because their jet inlet radius is half of ours.

However,  it is interesting to note that results of WB12 converge closer to ours, for the same cloud sizes, in spite of the different values of mean density of the clouds in the simulations. Hence, it appears that beyond a certain density, the core of the dense clouds remain unaffected, but feedback efficiency or coupling strongly depends on the spacing between clouds.

\subsubsection{Quantification of turbulence in the disk}    \label{sec:compression_ratio}

\begin{figure}
	\includegraphics[scale=0.45]{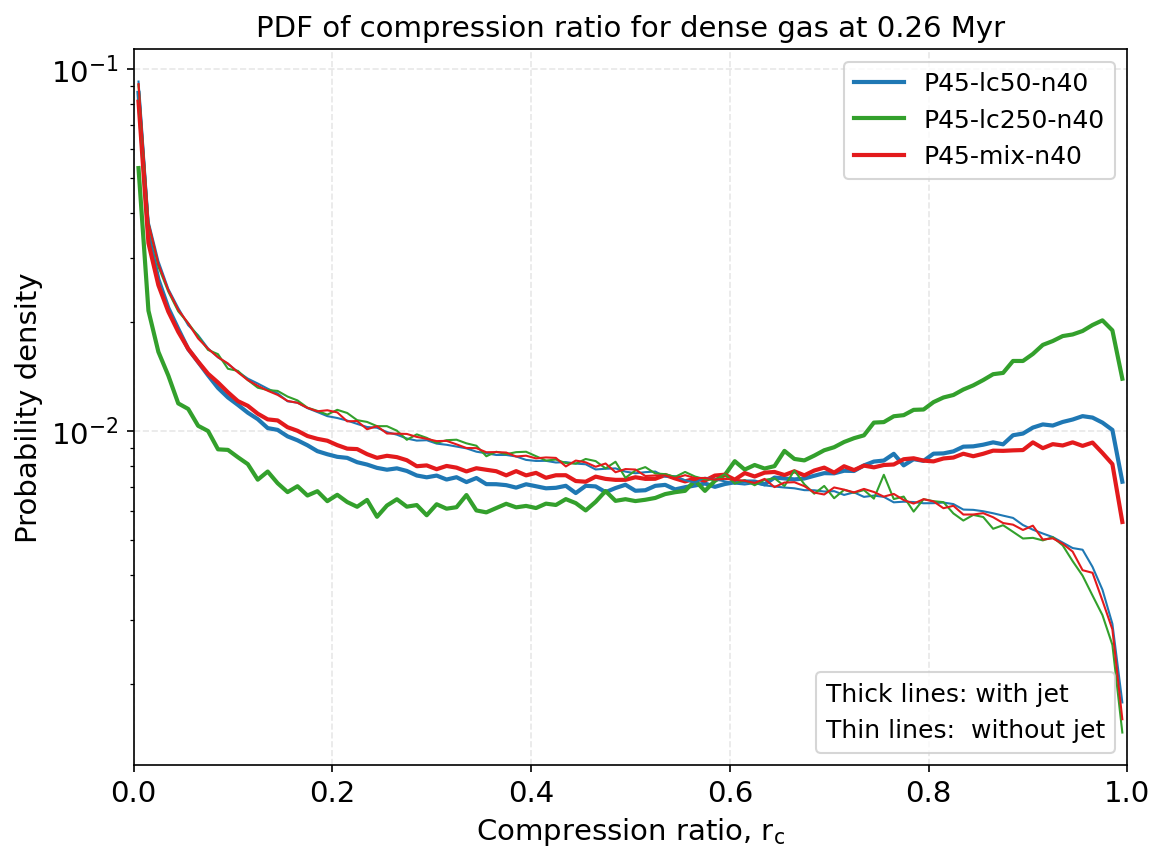}
    \caption{PDFs of compression ratio $r_{c}$ of a dense gas (defined as $n>0.1\,\,\rm cm^{-3}$) at 0.26 Myr for jetted (thick lines), and no jet (thin lines) simulations for different cloud configurations.}
    \label{fig:compression_ratio_PDF_compare_jet_nojet_cloudsize}
\end{figure}

As we have seen in the last subsection, jet-driven flows inject turbulence into the interstellar medium, making it essential to quantify the relative contributions of solenoidal ($\nabla\cdot\mathbf{v} = 0$) and compressible ($\nabla \times \mathbf{v} = 0$) modes to the turbulent velocity field. Fig.~(\ref{fig:vorticity_compression_ratio_disk_t1}) presents slices of the vorticity component $\omega_y$ (upper panel) and the compression ratio $r_{c} = |\nabla\cdot\mathbf{v}|^{2}/(|\nabla\cdot\mathbf{v}|^{2} + |\nabla \times \mathbf{v}|^{2})$ (lower panel) for the dense ISM at $t = 0.26$ Myr across different cloud configurations. We find that both the $l_{\rm cmax} = 50$~pc and mixed cases are predominantly solenoidal, whereas the $l_{\rm cmax} = 250$~pc case exhibits a stronger compressible contribution. This is further quantified by the probability density functions (PDFs) of the compression ratio shown in Fig.~(\ref{fig:compression_ratio_PDF_compare_jet_nojet_cloudsize}). In the absence of a jet (denoted by thin lines), all cloud configurations exhibit nearly identical PDFs, with a mean compression ratio of $\sim0.35$, indicating that the turbulence is predominantly solenoidal in nature. The injection of the jet, however, drives compressible outflows, with the $l_{\rm cmax} = 250$~pc case showing the highest mean compression ratio of $\sim0.54$. The mixed and $l_{\rm cmax} = 50$~pc cases yield mean compression ratios of $\sim0.41$ and $\sim0.42$, respectively, indicating comparable behaviour. 

The above results can be understood in the context of the dependence of the ``flood and channel'' phase of jet-ISM interaction \citep[see the right panel of Figure 1 in][]{mukherjee_2025_review} on the ISM structure. In the $l_{\rm cmax} = 50$~pc and mixed configurations, the dense, closely spaced clouds fragment the jet into multiple bifurcated channels, effectively reducing the power available per channel. As a consequence, the jet energy is dissipated more through shear and turbulent mixing rather than direct compression, resulting in predominantly solenoidal flows. In contrast, the $l_{\rm cmax} = 250$~pc case, characterised by large inter-cloud voids, allows the jet to propagate with minimal fragmentation. The full ram pressure of the jet thus acts directly on the massive clouds, leading to strong compression of the cloud bodies and even their dense cores. This efficient compression is reflected in the significantly higher compression ratio observed for this configuration. In addition to this, upon inspecting the time evolution of the compression ratio PDFs, we observe convergence of the PDFs across all cloud configurations at later times. This suggests that the turbulent state becomes increasingly independent of the initial cloud morphology as the jet-ISM interaction evolves.

\subsection{Effect of jet power $P_{\rm jet}$}\label{sec.jetPower}

\begin{figure*}
	\includegraphics[scale=0.50]{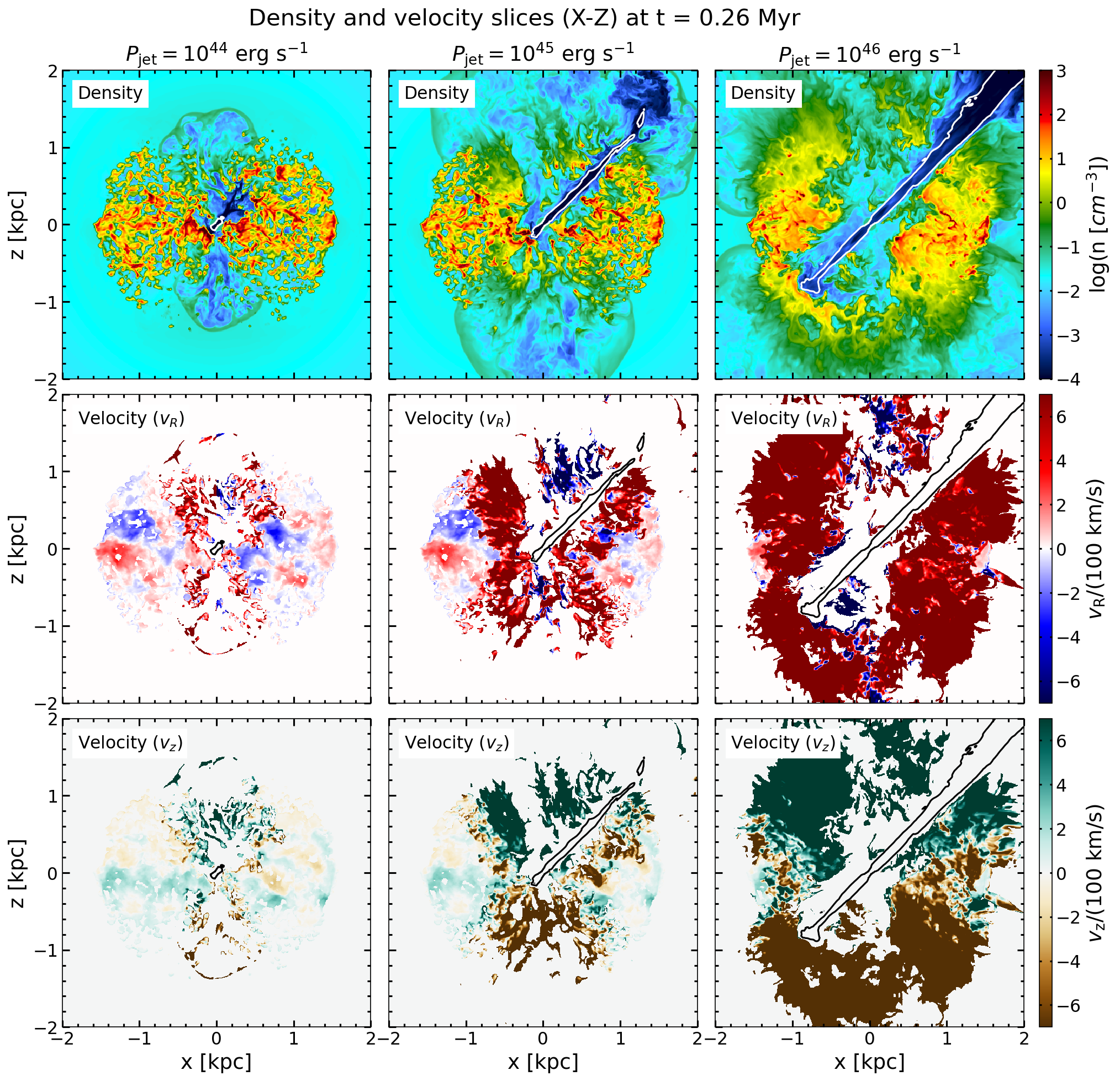}
    \caption{Density and velocity slices in the X-Z plane at 0.26 Myr for the simulations with $P_{\rm jet} = 10^{44}\,\,\rm erg\,\,s^{-1}$ (\textit{left}), $P_{\rm jet} = 10^{45}\,\,\rm erg\,\,s^{-1}$ (\textit{middle}), and $P_{\rm jet} = 10^{46}\,\,\rm erg\,\,s^{-1}$ (\textit{right}). Top: density ($\rm log\,\,n[cm^{-3}]$), middle and bottom: cylindrical radial velocity ($v_{\rm R}$), and vertical velocity ($v_{\rm z}$) normalized to 100 $\rm km\,\,s^{-1}$ for the dense gas (defined here as $n > 0.1\,\,{\rm cm^{-3}}$). The white and black contours in density and velocity profiles respectively, denote relativistic speeds with $\beta = 0.7$.
    }
    \label{fig:rho_vel_P44_P45_P46_xz_t1}
\end{figure*}

To explore how the strength of the jet impacts its interaction with the ISM, we carried out simulations using a mixed cloud configuration with three different jet powers: $10^{44}$, $10^{45}$, and $10^{46}\,{\rm erg\,\,s^{-1}}$ (Simulations P44-mix-n40, P45-mix-n40, P46-mix-n40, respectively; see Table~\ref{tab:simulation_list}). In the highest power case ($10^{46}\,{\rm erg\,\,s^{-1}}$), we additionally investigate the impact of discontinuous jet activity by switching off the jet after $\sim$0.85 Myr of continuous injection. This setup enables us to understand how the system evolves once the jet shuts down, particularly in terms of how long the induced turbulence and outflows persist in the absence of sustained energy input. These simulations allow us to assess how the energy input from the jet influences the degree of turbulence, gas dispersal, and breakout dynamics. The $X-Z$ slices of density and velocity at $t = 0.26$ Myr are presented in Fig.~(\ref{fig:rho_vel_P44_P45_P46_xz_t1}).

 \begin{itemize}
     \item \emph{\underline{Higher power jet ($P_{\rm jet} = 10^{46}\,\,\rm erg\,\,s^{-1}$)}: }
The higher power, $P_{\rm jet} = 10^{46}\,\,\rm erg\,\,s^{-1}$, breaks out of the ISM disk earlier and more easily, due to the large momentum it imparts to the ambient medium. It produces the strongest large-scale outflows, with mean cylindrical radial velocities exceeding $850\,\, \mbox{km s}^{-1}$. The total outflowing mass is $M_{v_R>280} \approx 1.3\times10^{9}\,\,M_{\odot}$, at a rate of $\sim 2320\,\,M_{\odot}\,\,\rm yr^{-1}$. Such strong outflows also result in very high values of mass weighted velocity dispersion of the outflowing gas, as shown in the bottom panel of Fig.~(\ref{fig:vel_dispersion_KE}). We find that the highest dispersions are achieved in the case of $P_{\rm jet} = 10^{46}\,\,\rm erg\,\,s^{-1}$, with $\sigma_{\rm R} \sim 850\,\,\rm km\,\,s^{-1}$ and $\sigma_{\rm z} \sim 1050\,\,\rm km\,\,s^{-1}$. The rate of kinetic energy transfer to the ISM around the jet breakout time is also the most efficient in this case, attaining  values up to 50\% of the jet power. However, this rapidly decreases  to  values even lower than that of the moderate and lower power jets,  due to the earlier decoupling of the jet from the ISM (Fig.~\ref{fig:KE_jet_fraction}, bottom panel).

Interestingly, in the case of $P_{\rm jet} = 10^{46}\,\,\rm erg\,\,s^{-1}$, the evolution of $\sigma_{\rm R}$ exhibits a distinct double-peaked structure, which can be attributed to the sequential breakout of the forward and counter jet lobes—first around $t \sim 0.1$ Myr and then at $t \sim 0.3$ Myr (see Sec.~\ref{sec:jet_confinement}). Following these breakout events, the velocity dispersion drops sharply, indicating a decoupling of the jet from the dense ISM disk. Furthermore, since the jet is turned off at $t \sim 0.85$ Myr in this simulation, the system experiences a more rapid decline in velocity dispersion as compared to the lower power jets, consistent with the cessation of energy injection and the gradual decay of turbulence in the ISM.

\item {\underline{Moderate to low power jets ($P_{\rm jet} \leq 10^{45}\,\,\rm erg\,\,s^{-1}$):}}
In contrast, jets with $P_{\rm jet} = 10^{45}$ and $10^{44}\,\,\rm erg\,\,s^{-1}$ break out later and exhibit progressively less gas dispersal. The outflows are localised to the vicinity of the jet-beam and with progressively lower speeds for lower jet power. The mean cylindrical radial velocity are $\overline{v_R} \lesssim 90 \mbox{ km s}^{-1}$ for $P_{\rm jet} = 10^{45}\,\,\rm erg\,\,s^{-1}$ (with $M_{v_R>280}\approx 2.11 \times 10^{8}\,\,M_{\odot}$, and outflow rate $\sim 508\,\,M_{\odot}\,\,\rm yr^{-1}$) and $\overline{v_{R}} \lesssim 10 \mbox{ km s}^{-1}$ for $P_{\rm jet} = 10^{44}\,\,\rm erg\,\,s^{-1}$ (with $M_{v_R>280}\approx 2.16 \times 10^{7}\,\,M_{\odot}$, and outflow rate $\sim 403\,\,M_{\odot}\,\,\rm yr^{-1}$). The lowest power jet in fact shows very small scale outflows, with the larger portion of the disc relatively unperturbed. For $P_{\rm jet} = 10^{45}\,\,\rm erg\,\,s^{-1}$, the dispersions are more moderate, around $250\,\,\rm km\,\,s^{-1}$ in both components, while the weakest jet ($10^{44}\,\,\rm erg\,\,s^{-1}$) leads to the lowest values, $\sim 150\,\,\rm km\,\,s^{-1}$. It is interesting to see that in case of the lower power jet, the kinetic energy transfer rate to the ISM remains positive for a significant amount of time and exhibits highest fraction of jet power among the three simulations (Fig.~\ref{fig:KE_jet_fraction}, bottom panel). This is because the lower power jet is confined within the disk for significant amount of time throughout the course of our simulations.
Such low to moderate impact of lesser power jets is consistent with earlier results of jet-disk simulations \cite{tanner_and_weaver_2022}, as well as simulations of \citet{wagner_2012} and \citet{mukherjee_2016} exploring jets in spherically distributed ISM.

\end{itemize}

The right panel of Fig.~(\ref{fig:outflow_vR_lmax_power}) further strengthens the conclusion that higher power jets produce more energetic outflows, and vice versa. Here we plot the maximum mass-weighted mean radial velocity ($\overline{v_{\rm R}}$) as a function of jet power (filled red circles). The values from the simulations $\rm C^{\prime}$, $\rm E^{\prime}$ and $\rm F^{\prime}$ of \cite{wagner_2012} are also shown for comparison (filled black triangles). We find that max($\overline{v_{\rm R}}$) = 52, 129, and 715 $\rm km\,s^{-1}$ for our simulations with jet powers of $10^{44}$, $10^{45}$, and $10^{46}$ $\rm erg\,s^{-1}$, respectively. The corresponding values from Wagner+12 are larger by factors of $\sim$9, 5, and 2, respectively, for the reasons discussed in Sec.~(\ref{sec:veldisp}).

\subsubsection{Accretion rates}

\begin{figure}
	\includegraphics[scale=0.50]{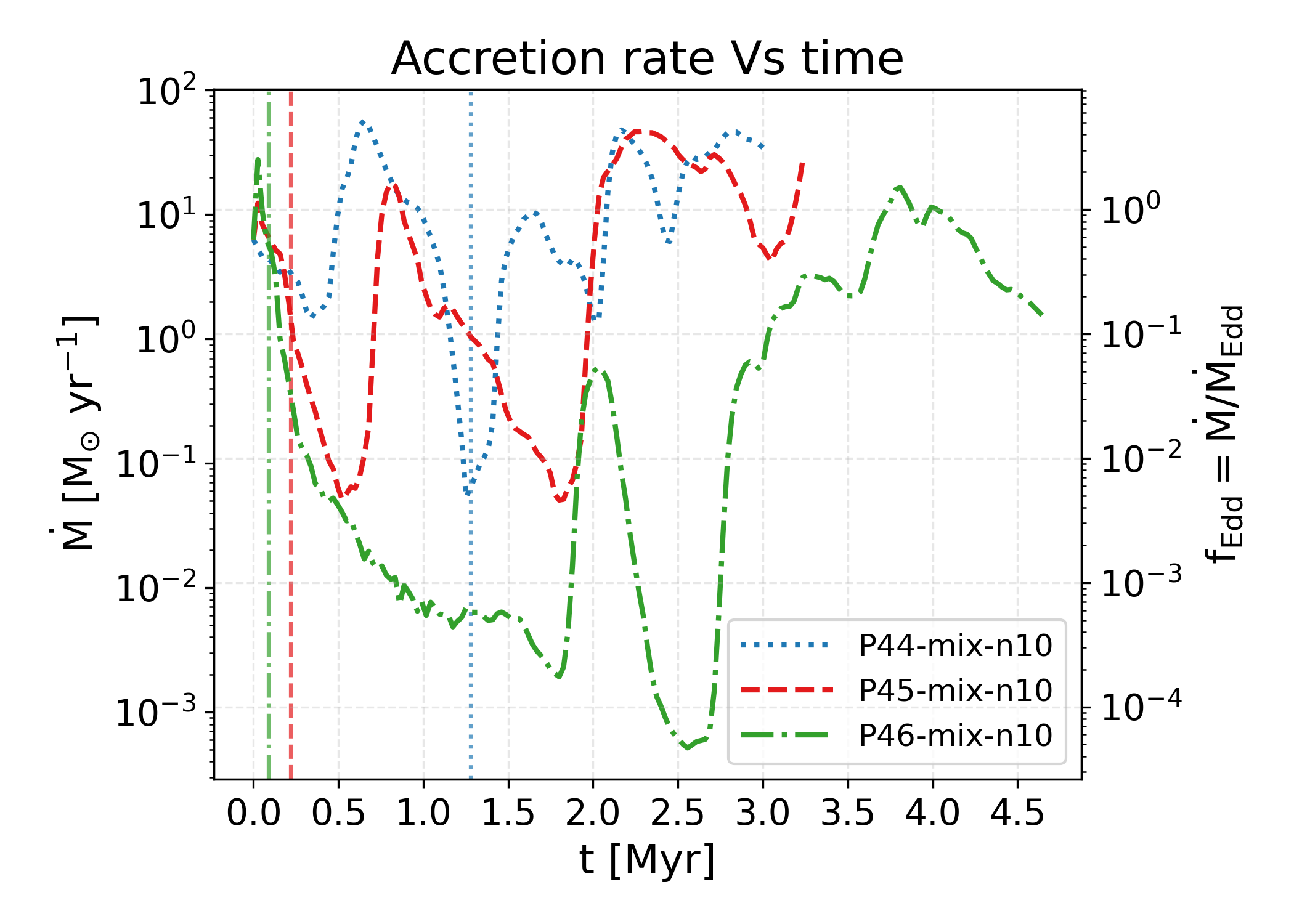}
    \caption{Accretion rate and Eddington ratio as a function of time.}
    \label{fig:accretion_rate_P44_P45_P46}
\end{figure}

To understand the impact of the jet on inflow of matter close to the nuclear region ($\sim 100\,\,\rm pc$), we study the time evolution of the accretion rate in post-process. Accretion rates are evaluated by computing the inward mass flux at a radius of 100~pc. Fig.~(\ref{fig:accretion_rate_P44_P45_P46}) shows the accretion rate profiles for simulations with different jet powers. The accretion rate (in $\rm M_{\odot}\,yr^{-1}$) and the corresponding Eddington ratio ($\rm f_{Edd} = \dot{M}/\dot{M}_{Edd}$) are shown on the same plot.\footnote{The Eddington accretion rate is calculated using $$\dot{M}_{\rm Edd} = \frac{L_{\rm Edd}}{\eta c^{2}} = \frac{4 \pi G M_{\rm BH} m_{\rm H}}{\eta \sigma_{\rm T} c},$$ with $M_{\rm BH} = 5 \times 10^{8} M_{\odot}$ (representative of 3C\,326\,N, \citealt{3c326_ogle_2007}), and $\eta = 0.1$.} As mentioned earlier, we note that, unlike the other simulations, the jet with power $10^{46}\,\,\rm erg\,s^{-1}$ is switched off at $t \sim 0.85$~Myr. For the simulations where the jet remains active, the accretion rates are highly variable, often reaching $f_{\rm Edd}\gtrsim 10\%$. This would imply that significant variations in the jet powers are expected within the time-scale of $\sim 1-2$ Myr. Also, even though the moderately powerful jet of $P_{\rm jet} =10^{45}\mbox{erg s}^{-1}$ drives outflows, there is still significant inward radial inflows which can potentially keep powering the jet and in fact change its power.  This suggests that firstly, assuming a constant jet power throughout the simulation is an oversimplification, and secondly, moderately powerful jets do not imply immediate cessation of accretion. Intriguingly, even when the jet is turned off (in the highest-power case), the accretion flow re-forms within 2--3~Myr and reaches $\rm f_{Edd} \gtrsim 1$. Thus self-regulation within a time scale of a few Myr is expected, as has also been implied in a some other previous simulations \citep{cielo_2018,clavijo24a}. Future studies should include accretion-regulated feedback \citep{husko26a}  for a more self-consistent investigation of jet-ISM feedback (Gogia et al. in prep).

% \begin{figure*}
% 	% To include a figure from a file named example.*
% 	% Allowable file formats are eps or ps if compiling using latex
% 	% or pdf, png, jpg if compiling using pdflatex
% 	\includegraphics[scale=0.5]{images/vel_cyl_P44_P45_P46_xz.png}
%     \caption{X-Z plane slices of cylindrical radial velocity ($v_{\rm R}$, \textit{top}) and the vertical velocity ($v_{\rm z}$, \textit{bottom}), normalized to 100 $\rm km\,\,s^{-1}$ for the dense gas (defined here as $n > 0.1\,\,{\rm cm^{-3}}$) at $t = 0.26$ Myr. \textit{Left to right} shows variation in jet power. The black contours denote $\beta = 0.7$.}
%     \label{fig:vel_cyl_P44_P45_P46_t1}
% \end{figure*}

\subsection{Effect of disk central density $n_{\rm w0}$}

\begin{figure*}
	\includegraphics[scale=0.5]{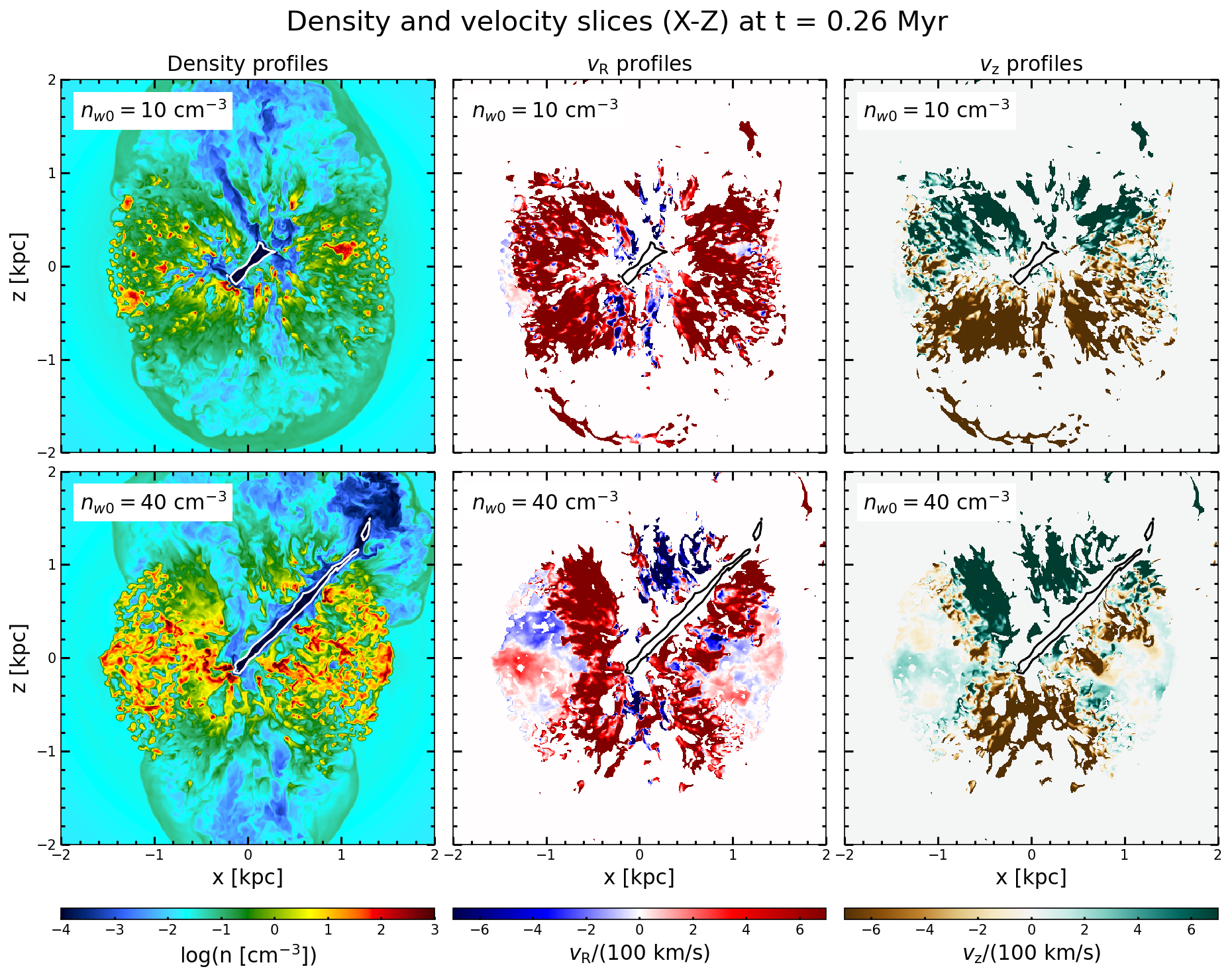}
    \caption{\textit{Left to right}: Density ($\rm log\,\,n[cm^{-3}]$), radial velocity ($v_{\rm R}$), and vertical velocity ($v_{\rm z}$) profiles at 0.26 Myr in the X-Z plane for the simulations P45-mix-n10 (\textit{top}), P45-mix-n40 (\textit{bottom}). The white contours denote $\beta = 0.7$.}
    \label{fig:vel_cyl_mixed_n10_n40_xz_t1}
\end{figure*}

\begin{figure}
	\includegraphics[scale=0.50]{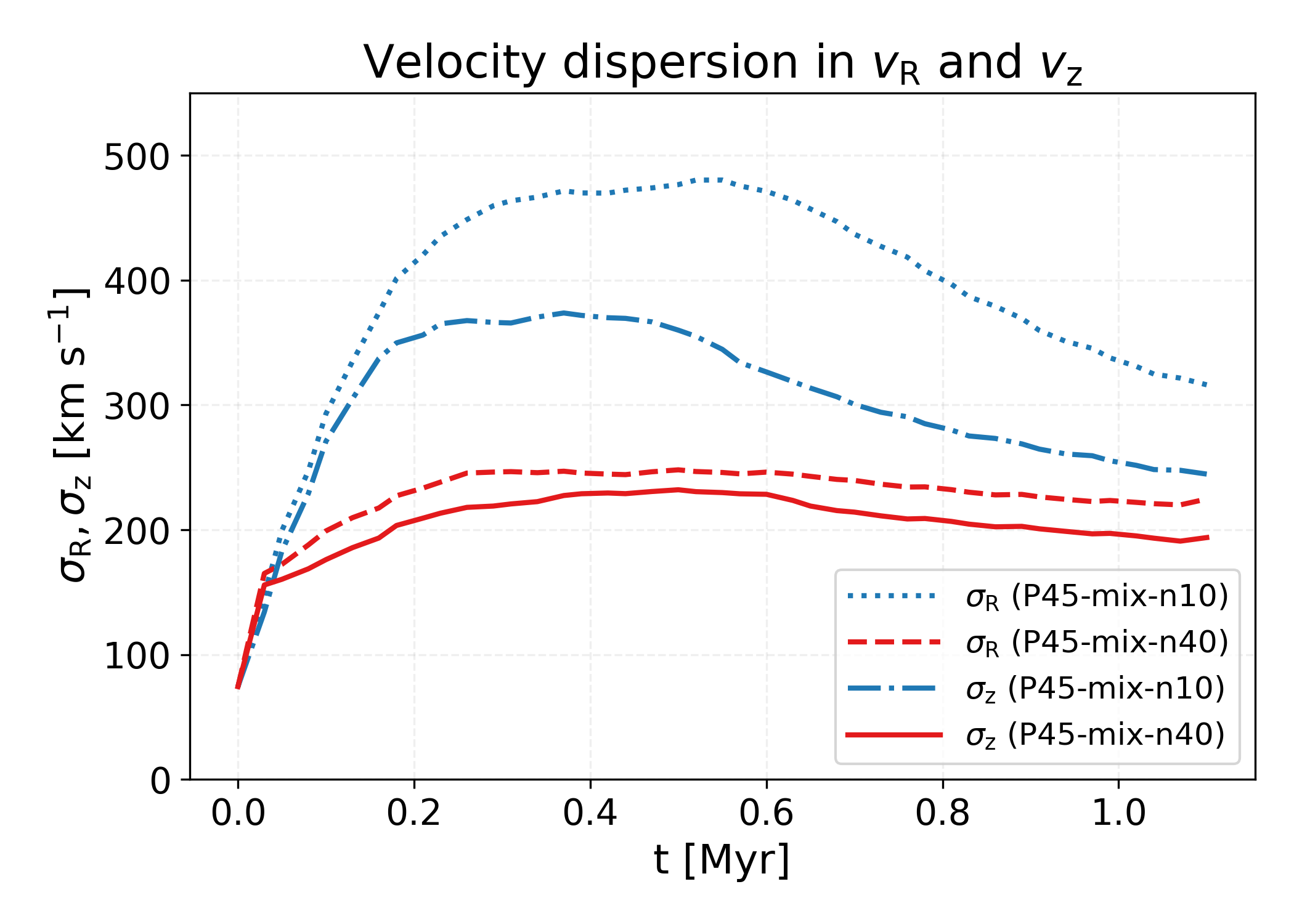}
    \caption{Mass-weighted velocity dispersion of dense gas (defined here as $n > 0.1\,\,\rm cm^{-3}$)}
    \label{fig:vel_dispersion_cyl_n10_n40_mixed}
\end{figure}

To examine the impact of different disk central densities ($n_{\rm w0}$) on jet–ISM interaction, we have also run a simulation with $n_{\rm w0} = 10\,\,\rm cm^{-3}$ using the mixed cloud configuration (P45-mix-n10; see Table~\ref{tab:simulation_list}), in addition to the other simulations which adopt $n_{\rm w0} = 40\,\,\rm cm^{-3}$. Fig.~(\ref{fig:vel_cyl_mixed_n10_n40_xz_t1}) shows the density (\textit{left}) and velocity structures (\textit{middle and right}) for the simulations having mean central densities $n_{\rm w0} = 10\,\,\rm cm^{-3}$ (\textit{top}), and $n_{\rm w0} = 40\,\,\rm cm^{-3}$ (\textit{bottom}) at the epoch $t = 0.26$ Myr. We observe that the ISM with low density ($n_{\rm w0} = 10\,\,\rm cm^{-3}$) is more stirred up by the impact of a jet when compared to its higher density counterpart ($n_{\rm w0} = 40\,\,\rm cm^{-3}$), and thus produces more powerful, diffused outflows. 

The above findings are also reflected in the plots of velocity dispersion presented in Fig.~(\ref{fig:vel_dispersion_cyl_n10_n40_mixed}). The simulation with the lower central density exhibits significantly higher velocity dispersions, with peak values nearly twice those of the higher-density case. This enhancement in velocity dispersion at lower $n_{\rm w0}$ can be understood in terms of the resistance offered by the ambient medium. A lower-density ISM presents less inertia to the propagating jet, allowing it to penetrate more easily and stir up the gas more efficiently. In contrast, a denser medium absorbs more of the jet’s energy locally and resists bulk acceleration, thereby suppressing the development of large-scale turbulent motions. As a result, the energy transfer from the jet to the ISM is more widespread and kinematically effective in the low-density case, leading to enhanced gas agitation and greater velocity dispersions,
again showing consistency with the simulations of \cite{wagner_2012, mukherjee_2018}. %{\color{teal} (Note to Mayur: ** TO DO ** Again add more stuff to compare with Wagner.)}

\subsection{Jet Propagation and Confinement Timescales}     \label{sec:jet_confinement}

\begin{figure}
	\includegraphics[scale=0.55]{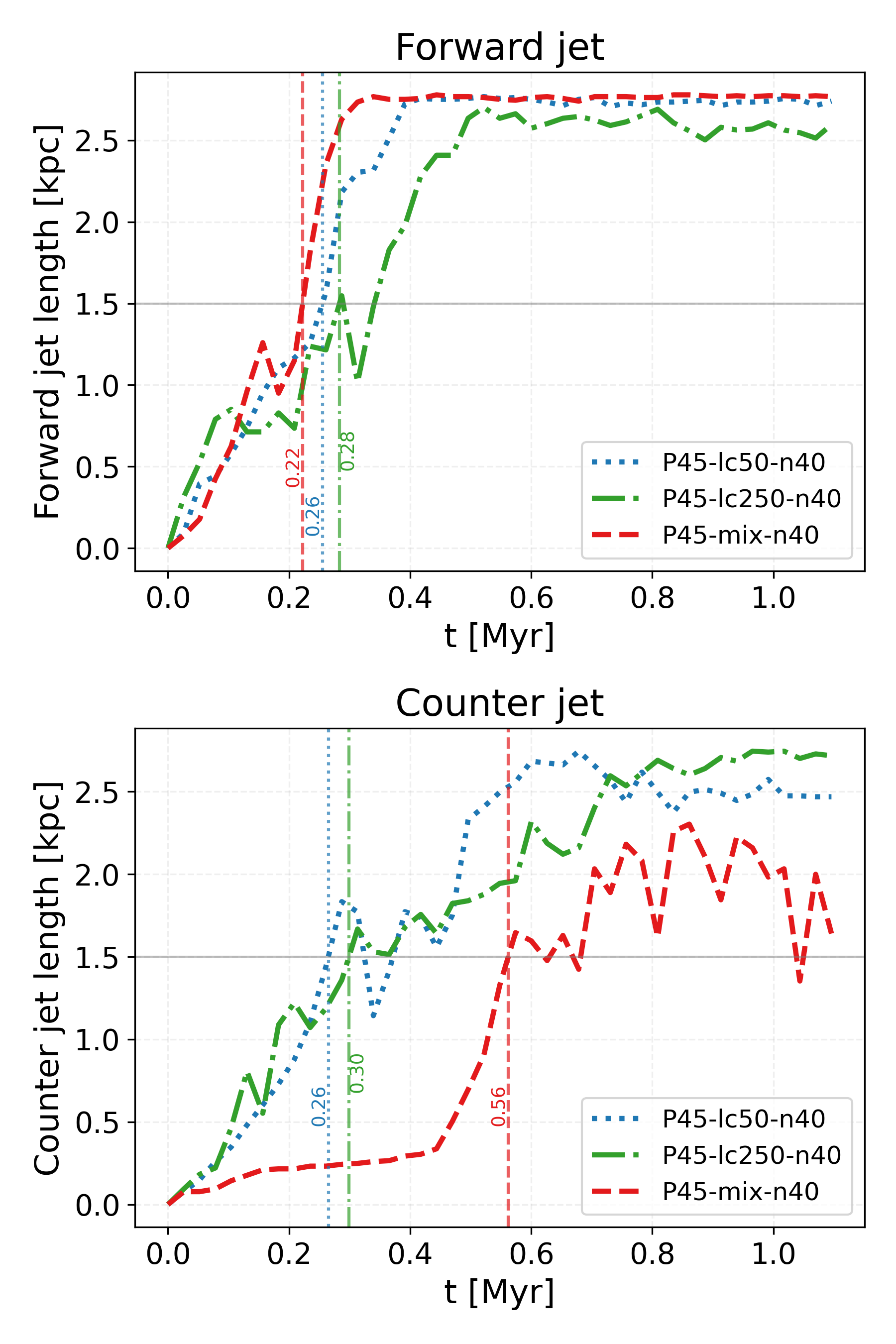}
    \caption{Evolution of the lengths of forward jet (\textit{top}) and counter jet (\textit{bottom}) with time (here, jet is defined as the material with velocity $\beta > 0.4$) for different cloud configurations. Jet breakout times are denoted by corresponding vertical lines.}
    \label{fig:jet_length_100_500_mixed}
\end{figure}

The jet length and its age serve as indicators of the degree to which the ISM confines the jet over time.
%The evolution of jet length serves as a key indicator of the degree to which the ISM confines the jet over time. 
Fig.~(\ref{fig:jet_length_100_500_mixed}) shows the temporal evolution of the lengths of the forward jet (the one which is propagating into the upper half of the grid; top panel) and the counter jet (the one which is propagating into the lower half; bottom panel), for simulations with different cloud configurations. Jet material is defined as gas with velocity $\beta > 0.4$, allowing us to isolate the high-velocity outflowing components.

We define the jet as having broken out of the disk once it exceeds a length of approximately 1.5 kpc (since the radius of the disk is 1.5 kpc.) Also, we cannot track the jet once it goes out of the computational domain. In such a case, we see the saturation of the jet-length to $\sim 2.8\,\,\rm kpc$, which we observe in case of forward jets of $l_{\rm cmax} = 50\,\,\rm pc$ and mixed configurations (top panel, Fig.~\ref{fig:jet_length_100_500_mixed}). To first order, we expect that the jet breakout should be delayed in simulations where the jet encounters larger clouds, owing to greater inertia. With the same logic, configurations dominated by smaller clouds should allow for relatively earlier breakout, as the jet more easily disrupts and navigates through the less massive clumps. So we expect a clear trend from studying the evolution of jet length. But the situation is more involved than this. The relative location of clouds in the close vicinity of the jet path also affects the evolution. Hence, we see the complex evolution of the jet length in Fig.~(\ref{fig:jet_length_100_500_mixed}), wherein the forward jet in $l_{\rm cmax} = 250\,\,\rm pc$ case has the largest breakout time ($\sim 0.28\,\,\rm Myr$) compared to the other two cases, but the mixed configuration shows an earlier breakout ($\sim 0.22\,\,\rm Myr$) than that of $l_{\rm cmax} = 50\,\,\rm pc$ ($\sim 0.26\,\,\rm Myr$).

Moreover, the evolution of both forward and counter jets need not be similar. This is evident in the bottom panel of Fig.~(\ref{fig:jet_length_100_500_mixed}). In the mixed cloud configuration, we indeed observe such a scenario where the forward jet encounters small clouds and the counter jet interacts with large clouds. As a result, the counter jet experiences prolonged confinement and a significantly delayed breakout time compared to the other two cases. This introduces asymmetry in the arm-lengths of the jet. This has also been observed in extragalactic radio sources \citep{mccarthy_1993, nesvadba_2008} and a possible explanation given by \cite{gaibler_2011} via 3D hydrodynamic simulations resorted to the inhomogeneities in the ISM, consistent with our study. These findings emphasize that the scale of clouds and their relative arrangement with the jet path play a critical role in setting the confinement timescale of the jet and hence the efficiency with which the jet can transfer energy and momentum to the ISM.
%When the jet interacts with the larger clouds present in this distribution, it experiences prolonged confinement similar to that seen in the $\lambda_{\rm max} = 500\,\,\rm pc$ case. As a result, the breakout time in the mixed case can be significantly delayed, depending on the specific cloud structures along the jet’s path. These findings emphasize that the presence of large clouds—regardless of whether they dominate the distribution—plays a critical role in setting the confinement timescale of the jet and hence the efficiency with which the jet can transfer energy and momentum to the ISM.

\begin{figure}
	\includegraphics[scale=0.55]{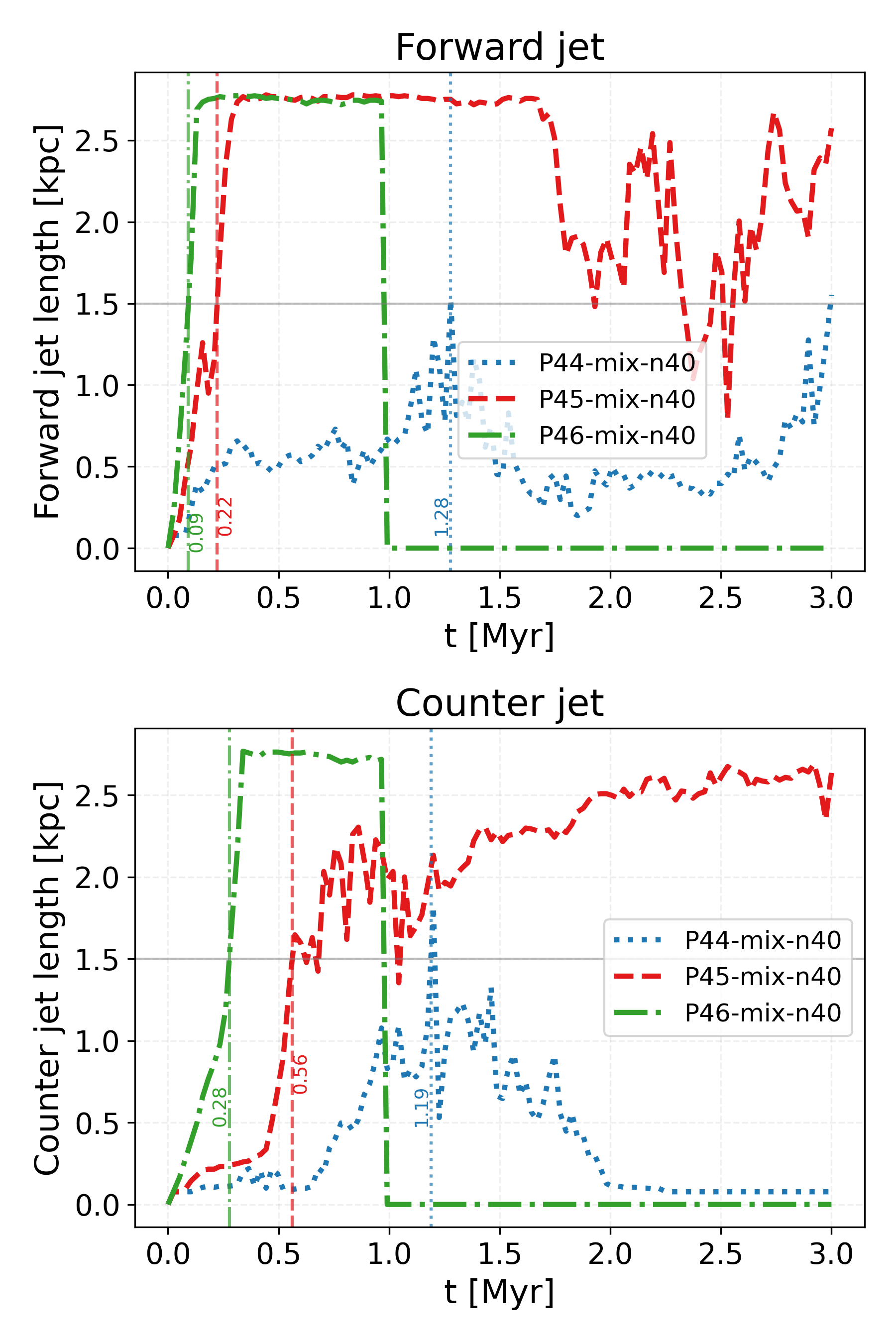}
    \caption{Evolution of the lengths of forward jet (\textit{top}) and counter jet (\textit{bottom}) with time (here, jet is defined as the material with velocity $\beta > 0.4$) for different jet powers. Jet breakout times are denoted by corresponding vertical lines.}
    \label{fig:jet_length_P44_P45_P46}
\end{figure}

In addition to the influence of cloud configurations, the intrinsic power of the jet plays a major role in determining its propagation characteristics and breakout timescales. Fig.~(\ref{fig:jet_length_P44_P45_P46}) shows the temporal evolution of the lengths of the forward (top panel) and counter jets (bottom panel) for simulations with jet powers $P_{\rm jet} = 10^{44}$, $10^{45}$, and $10^{46}\,\,\rm erg\,\,s^{-1}$, all carried out using the same mixed cloud configuration. As expected, higher power jets propagate more efficiently through the ISM in the adopted range of gas density. The most powerful jet ($10^{46}\,\,\rm erg\,\,s^{-1}$) rapidly pierces the disk and breaks out in less than $\sim0.1$ Myr. However, since this jet was deliberately turned off at $t \sim 0.85$ Myr in the simulation to study intermittent activity, we observe a sharp decline in jet length around that time, eventually dropping to zero as the high-speed material disperses and no new jet material is injected.
In contrast, the lowest power jet ($10^{44}\,\,\rm erg\,\,s^{-1}$) remains confined within the disk throughout the entire simulation runtime. It lacks sufficient momentum to overcome the resistance posed by the clumpy ISM, resulting in continuous energy deposition within the disk and the absence of a true breakout event.
The intermediate case ($10^{45}\,\,\rm erg\,\,s^{-1}$) exhibits delayed breakout compared to the most powerful jet, but it still manages to eventually escape the disk environment. This highlights how increasing jet power not only accelerates the breakout process but also reduces the degree of interaction with the ISM. In contrast, weaker jets remain trapped longer and couple more efficiently with the surrounding gas, enhancing turbulence and feedback within the disk.

%{\color{teal} (Note to Mayur: Discuss this more quantitatively using Mukherjee's earlier works on jet confinement.)}

\subsection{Evolution of the phase diagram}     \label{sec:phase_diagram}

\begin{figure*}
	\includegraphics[scale=0.55]{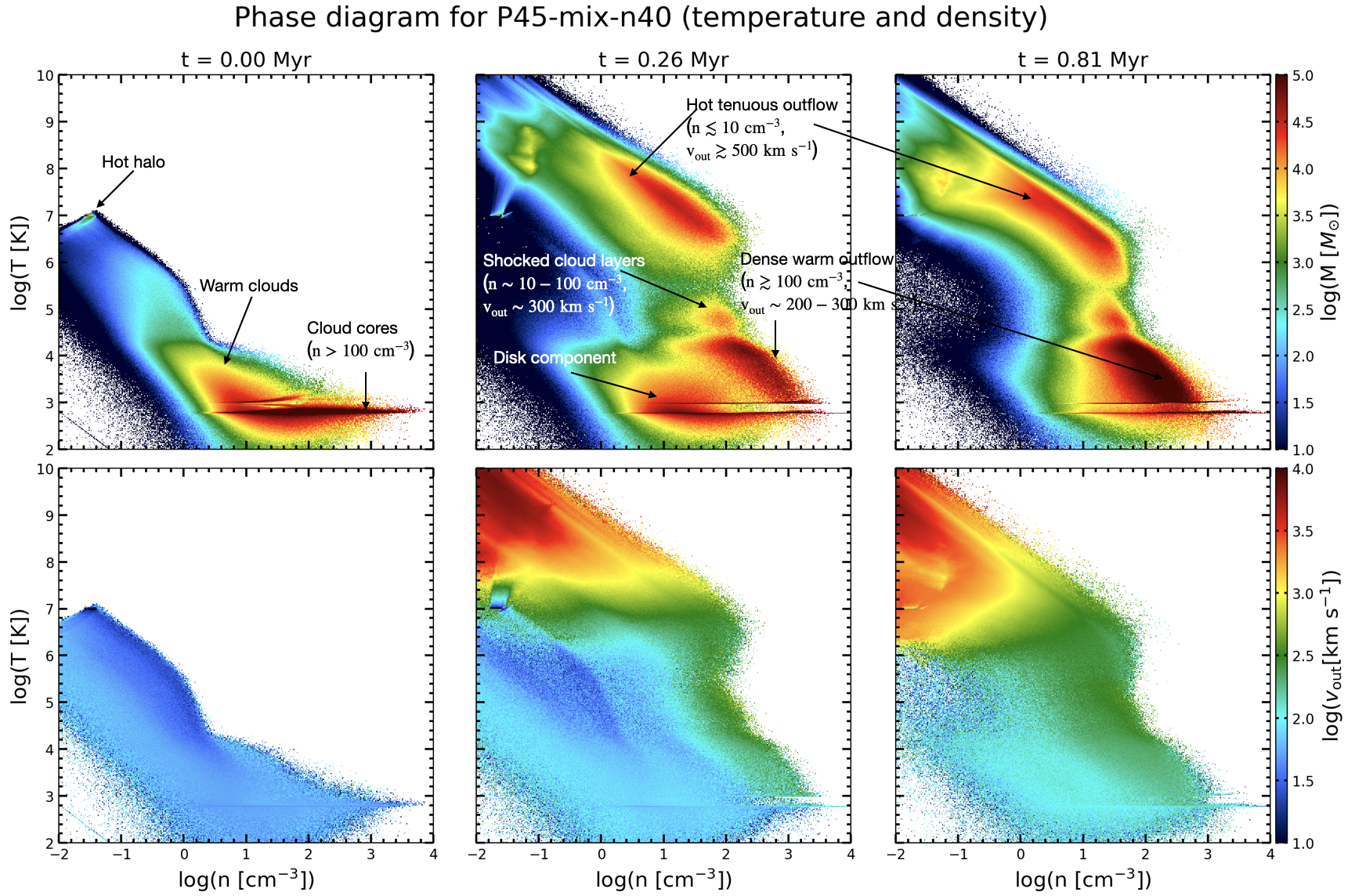}
    \caption{Evolution of phase diagram with time}
    \label{fig:phase_space}
\end{figure*}

Jet-ISM interaction gives rise to multiple coexisting gas phases, which manifest through different signatures in multi-wavelength observations. Phase diagrams are powerful tools to understand the multiphase nature of outflows, as they help us disentangle the thermodynamic and kinematic properties of the outflowing gas. Although the evolution of gas phases shown in the phase diagram depends
%evolution of phase diagram depends 
on the physical parameters under consideration--such as the nature of ISM distribution, jet power, and also the densities--the overall structure attributing to different gas phases remains similar. Hence, in this subsection, we focus on the evolution of gas phases
%evolution of phase diagram 
for a representative simulation P45-mix-n40, by exploring the space consisting of density ($n$), temperature ($T$), and outflow velocity ($v_{\rm out}$). Such an analysis is shown in Fig.~(\ref{fig:phase_space}), wherein the mass distribution (\textit{top}) and outflow velocity distribution (\textit{bottom}) are plotted in the two dimensional phase structure of temperature and gas density. Just before injection ($t=0$ Myr), the settled gas has 3 main phases--hot halo ($T\sim 10^{7}\,\,\rm K$ and $n\lesssim0.1\,\,\rm cm^{-3}$), dense cloud cores ($T\sim 10^{3}\,\,\rm K$ and $n>100\,\,\rm cm^{-3}$), and warm outer cloud layers ($T\sim 10^{3}-10^{4}\,\,\rm K$ and $n\sim 1-100\,\,\rm cm^{-3}$). During the jet-ISM interaction, we notice the following  distinct gas phases,  as also outlined in the review by \citealt{mukherjee_2025_review}:
\begin{itemize}
    \item \textit{Cloud cores and disk component}: These are the phases before jet injection (dense cloud cores and warm outer cloud layers), having velocities $v_{\rm out} \lesssim 100\,\,\rm km\,\,s^{-1}$ which retain their nature during jet-ISM interaction. The disk component gradually diminishes with time because the outer layers of the clouds get ablated by jet impact.
    
    \item \textit{Dense warm outflow}: We see the large collection of mass in the temperature range $T\sim 10^{3}-10^{4}\,\,\rm K$, outflow velocities ranging $v_{\rm out}\sim 200-300\,\,\rm km\,\,s^{-1}$, and having the density $n\gtrsim 100\,\,\rm cm^{-3}$. This corresponds to dense, shock-heated gas that has cooled and been accelerated by the jet; it represents the dominant mass in the outflow.
    
    \item \textit{Shocked cloud layers}: A small mass collection having temperatures $T\sim 10^{4}-10^{5}\,\,\rm K$, density $n\sim 10-100\,\,\rm cm^{-3}$, and accelerated to the velocities $v_{\rm out}\sim 300\,\,\rm km\,\,s^{-1}$ is discernible. This corresponds to the outer layers of clouds which are shocked to high temperatures by the jet interaction.
    
    \item \textit{Hot tenuous outflow}: This is the jet-driven phase of the gas having the highest outflow velocities ($v_{\rm out} \gtrsim 500\,\,\rm km\,\,s^{-1}$), very high temperatures ($T > 10^{6}\,\,\rm K$), and very low densities ($n\lesssim 10\,\,\rm cm^{-3}$). 
\end{itemize}

Together, these phases represent the transformation of ISM material processed by the jet --- from quiescent cloud cores, through shocked and ablated layers, to fast, tenuous outflow. Each phase carries distinct kinematic and thermal signatures that collectively trace the dynamical and thermal evolution of the gas as it is progressively entrained, accelerated, and heated by the jet–ISM interaction.

\section{Comparison of simulations with the jet-driven bubble in 3C\,326\,N}\label{sec.3C326}

Having explored the effects of different cloud configurations, jet powers, and cloud central densities on the dynamics of the jet--ISM interaction, we now turn our attention to connecting these results with observations. One of the primary motivations of this work is to understand the physical processes at play in radio galaxies exhibiting large-scale outflows and disturbed kinematics. In particular, the radio galaxy 3C\,326\,N shows compelling and unambiguous signatures of jet-driven feedback in its molecular and ionized gas phases. In the following section, we summarize key observational findings for 3C\,326\,N that are most relevant to our simulations, setting the stage for a direct comparison between the model predictions and observed properties.

\subsection{Observations of 3C\,326\,N}

%Observations done with the Spitzer/IRS revealed that 30\% of nearby ($z < 0.2$) powerful radio galaxies corresponding to 3CRR catalogue are ``${\rm H_{2}}$ luminous'' (Molecular Hydrogen Emission Galaxies--MOHEGs), showing bright, infrared line emission from ${\rm H_{2}}$ molecules, which does not appear to be associated with star formation \citep{3c326_ogle_2007, 3c326_ogle_2010, 3c326_nesvadba_2010, 3c326_nesvadba_2011}. The spectra are dominated by pure rotational lines of molecular hydrogen ($L(H_{2}) = 10^{40}-10^{43} \,\,{\rm erg\,\,s^{-1}}$), while the common star-formation tracers like bright IR continuum, mid-IR lines of [NeII] and [NeIII], and PAH lines are either very weak or absent. On strong physical grounds, \cite{3c326_ogle_2007} and \cite{3c326_nesvadba_2010, 3c326_nesvadba_2011} suggested that the bright $\rm H_{2}$ line emission is produced by the shocks created by the interaction of radio jet with the multiphase ISM.
3C\,326\,N ($z = 0.09$) is one of the best studied examples of the subset of $\rm H_{2}$ luminous radio galaxies \citep{3c326_ogle_2007, 3c326_nesvadba_2010, 3c326_nesvadba_2011}. It has one of the largest known radio sources, and has an FRII morphology having two Mpc-sized radio lobes \citep{3c326_willis_and_strom_1978, 3c326_rawlings_1990}. Spitzer/IRS observations of 3C\,326\,N by \cite{3c326_ogle_2007} revealed very strong $\rm H_{2}$ pure-rotational line emission in the $5-30\,\,\rm \mu m$ range, while the typical AGN features, as well as star formation tracers such as PAH emissions, were weak or absent. Star-formation rate being very low (SFR $\approx 0.087 \,\, \rm M_{\odot}\,\, yr^{-1}$), and a faint AGN ($L_{X} = 10^{40}\,\,\rm erg\,\,s^{-1}$) in 3C\,326\,N imply that neither of the above energy sources are powerful enough to heat the molecular gas \citep{3c326_nesvadba_2010}. Combining the Spitzer/IRS with IRAM CO(1--0) observations, \cite{3c326_nesvadba_2010} confirm the dominance of warm molecular gas of mass $\sim2\times10^{9} \,\,\rm M_{\odot}$ (which is believed to be heated by jet-driven shocks) in overall molecular gas budget in 3C\,326\,N ($\sim 2.5\times10^{9} \,\,\rm M_{\odot}$). All of this makes it a clear example of AGN feedback and a particularly good source for a detailed assessment of the impact of AGN feedback on the gas energetics and dynamics in their host galaxy.

\begin{figure*}
	\includegraphics[scale=0.363]{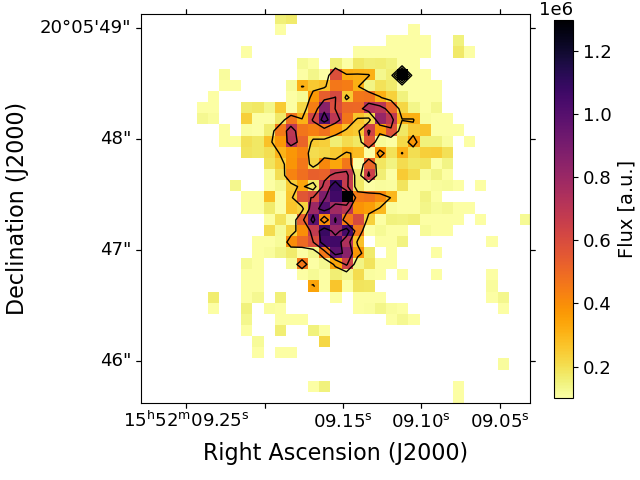}
    \includegraphics[scale=0.363]{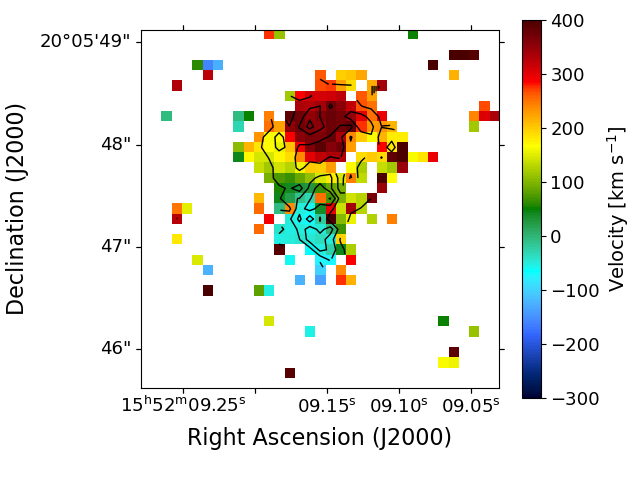}
    \includegraphics[scale=0.363]{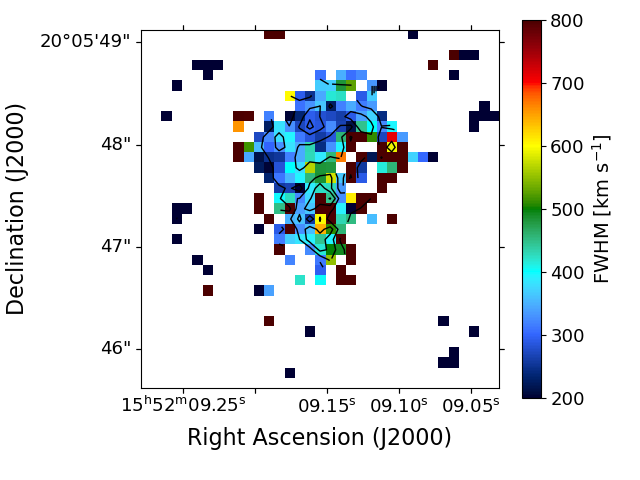}
    \caption{JWST/NIRSpec observations of 3C\,326\,N (systemic line component): $\rm H_{2}$ 1--0 S(3) emission morphology (\textit{left}), LOS velocities (\textit{centre}), and FWHM line widths (\textit{right})(similar to Fig.~3 of \citealt{3c326_leftley_2024}). The contours show emission line morphology. (Credits: Dr. James Leftley and Dr. Nicole Nesvadba)}
    \label{fig:3c326_observations}
\end{figure*}

More recently, this source was studied again by \cite{3c326_leftley_2024}, this time with higher sensitivity and higher resolution than the previously reported observations with the help of JWST/NIRSpec and MIRI imaging spectroscopy, on scales of order $\sim 100\,\,\rm pc$, commensurate with giant molecular cloud associations. They report bright ro-vibrational, and pure rotational lines of molecular hydrogen, which are the tracers of turbulence in the molecular gas in 3C\,326\,N. The key properties of 3C\,326\,N inferred from the observations which we want to reproduce in our simulations are:
\begin{enumerate}
    %\item Star-formation rate being very low (SFR $\approx 0.087 \,\, \rm M_{\odot}\,\, yr^{-1}$), and a faint AGN ($L_{X} = 10^{40}\,\,\rm erg\,\,s^{-1}$) in 3C\,326\,N imply that neither of the above energy sources are powerful enough to heat the molecular gas \citep{3c326_nesvadba_2010}.
    \item Emission line morphology confirms the presence of $\rm H_{2}$ molecular disk of size 3 kpc \citep{3c326_nesvadba_2011, 3c326_leftley_2024}.
    \item NIRSpec observations reveal a kpc-sized cavity in the northern part of the disk (Fig.~3 of \citealt{3c326_leftley_2024}), likely representing the possible scenario of an expanding, jet-driven bipolar bubble in the rotating disk.
    \item This bubble is delineated by three bright clumps in the far North (see Fig.~\ref{fig:3c326_observations}).
    \item The bubble is expanding with the maximal velocities of $380\,\,\rm km\,s^{-1}$.
    \item The FWHM of ro-vibrational lines are in the range 200--1000 $\rm km\,\,s^{-1}$.
\end{enumerate}

The systemic line component of 3C\,326\,N ($\rm H_{2}$ 1--0 S(3) emission) using JWST/NIRSPec is shown in Fig.~(\ref{fig:3c326_observations}), with emission flux, LOS velocity and FWHM line widths plotted in left, centre and right panels, respectively. We can see all the key properties of 3C\,326\,N listed above.

%\subsection{Simulations to explain the 3C\,326\,N observations}
\subsection{Connecting simulations with the observations of 3C\,326\,N}

To interpret the observed properties of 3C\,326\,N in the context of jet--ISM interactions, we now compare our simulation results with the observed characteristics of molecular hydrogen ($\rm H_{2}$) emission in this source. %Among the various simulations performed, the one with a jet power of $10^{45}\,\,\rm erg\,\,s^{-1}$ and a mixed cloud configuration (P45-mix-n40) most closely reproduces the salient features of 3C\,326\,N. The mixed cloud configuration refers to a combination of cloud distributions with $\lambda_{\rm max} = 100$ pc and $\lambda_{\rm max} = 500$ pc---representing small and large characteristic cloud sizes, respectively. This setup is motivated by the need to emulate a clumpy, multi-scale ISM comprising both compact and extended gas structures. Such a configuration enables a range of shock interactions, leading to a turbulent medium and extended outflows—consistent with the $\rm H_{2}$ emission morphology and kinematics seen in 3C\,326\,N. 
One of the striking observational features of 3C\,326\,N is the detection of a kiloparsec-scale cavity (or bubble) in the northern region of the disk, as revealed by high-resolution NIRSpec integral field spectroscopy \citep{3c326_leftley_2024}. This cavity, likely inflated by a jet-driven outflow, offers a compelling signature of AGN feedback shaping the multiphase ISM. To investigate whether a similar structure arises in our simulations, we generate synthetic emission maps and corresponding $W_{80}$ velocity widths through post-processing, following the method recently developed by \cite{meenakshi_2022}. This enables us to compare our simulations results directly with observations.
%Our research group has recently developed a method for producing synthetic line-emission and corresponding $W_{80}$ velocity width maps from simulation outputs \citep{meenakshi_2022}, enabling direct comparisons with observations. 
%Although our simulations do not model molecular $\rm H_{2}$ emission explicitly, we can qualitatively compare the observed $\rm H_{2}$ morphology with the emission from shocked gas in our models. 
Although our simulations do not model molecular $\rm H_{2}$ emission explicitly, we can compare the proxy emission from shocked gas in our models to the observed $\rm H_{2}$ morphology, since it is found that most of the $\rm H_{2}$ lines are predominantly shock dominated.

%Following the approach of \cite{meenakshi_2022}, who computed synthetic [O\,\textsc{iii}] maps by evaluating the volumetric emissivity $n^{2} \Lambda$, where $n\,\,[\rm cm^{-3}]$ is the gas number density and $\Lambda\,\,[\rm erg\,\,cm^{3}\,\,s^{-1}]$ is the cooling rate corresponding to [O\,\textsc{iii}] emission (extracted from \texttt{MAPPINGS V}; \citealt{sutherland_2018}), we adopt a simplified strategy here. 
The method devised by \cite{meenakshi_2022} consists of computing the volumetric emissivity $n^{2} \Lambda$, where $n\,\,[\rm cm^{-3}]$ is the gas number density and $\Lambda\,\,[\rm erg\,\,cm^{3}\,\,s^{-1}]$ is the cooling function corresponding to the required emission line.
Since we are focusing on the ro-vibrational $\rm H_{2}$ emission, which we cannot exactly calculate in the post-processing of our simulations, we approximate the emissivity as $\propto n^{2}$, by setting $\Lambda = \rm const = 1$, and then integrate $n^2$ along the line of sight to obtain the emission flux. Similar approximate measures of emission from shocked gas has also been used in earlier works \cite[e.g.][]{mukherjeeIC5063_2018} to investigate the predictions of the observable kinematics of the shocked dense gas in simulations. 

To trace the gas shocked to different conditions by the jet, we produce emission maps for various temperature intervals: $T=(1$--$5)\times10^{3}\,\,\rm K$, $T=(5$--$20)\times10^{3}\,\,\rm K$, $T=(20$--$50)\times10^{3}\,\,\rm K$, and $T=(50$--$500)\times10^{3}\,\,\rm K$. Since the emission morphology is also sensitive to the viewing geometry, we generate maps for multiple lines of sight (LOS) to determine which configuration best reproduces the observed structure. We find that the maps corresponding to a LOS with polar angle $\theta_{\rm I} = 150^{\circ}$ and azimuthal angle $\phi_{\rm I} = 90^{\circ}$ provide the closest morphological match to the observations of 3C\,326\,N. Here, $\theta_{\rm I}$ and $\phi_{\rm I}$ denote the inclination and azimuthal orientation of the observer, respectively. The subscript `I' indicates inclination with respect to the normal of the disk mid-plane.

\begin{figure}
	\includegraphics[scale=0.37]{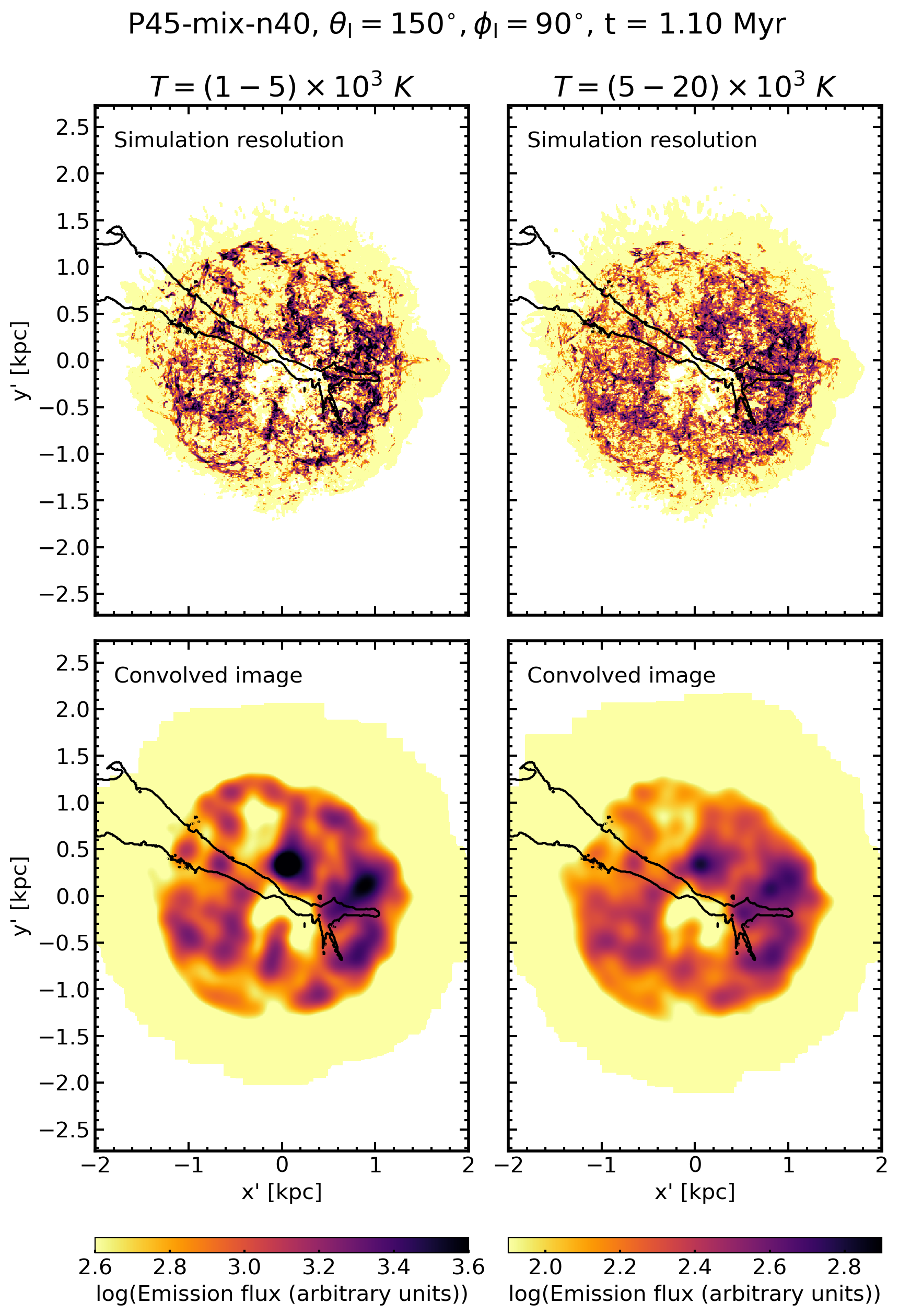}
    \caption{Synthetic emission maps for a fiducial simulation P45-mix-n40 at t = 1.1 Myr for the LOS with inclination angles $\theta_{\rm I} = 150^{\circ}$ and $\phi_{\rm I} = 90^{\circ}$. \textit{Left to right:} Shocked emission corresponding to gas in the temperature ranges $T = (1$--$5)\times10^{3}\,\,\rm K$, and $T = (5$--$20)\times10^{3}\,\,\rm K$, respectively. \textit{Top:} Unconvolved image at the resolution of the simulation. \textit{Bottom:} The image convolved with the PSF of $\rm 190\,\,pc \times190\,\,pc$. Black contours indicate the projected jet $\beta$ of 0.7, outlining the jet morphology.}
    \label{fig:loslum_150_90_all_P45_mixed_ct1_smooth}
\end{figure}

Fig.~(\ref{fig:loslum_150_90_all_P45_mixed_ct1_smooth}) presents synthetic emission maps from the fiducial simulation (P45-mix-n40) at t = 1.1 Myr, projected along this preferred LOS configuration. The left panel shows the shocked emission from gas in the temperature range $T = (1$--$5)\times10^{3}\,\,\rm K$, whereas the emission from the temperature range $(5$--$20)\times10^{3}\,\,\rm K$ is shown on the right. These two temperature slabs probe distinct phases of shocked ISM: the lower-temperature range traces the regions from where the ro-vibrational $\rm H_{2}$ emission is expected, whereas the higher-temperature range primarily corresponds to ionized gas emission. The top panels display images at the resolution of the simulation. To facilitate a meaningful comparison with the JWST observations, the simulated images are convolved with the JWST/NIRSpec point spread function (PSF), which is $\approx 0.11''$, corresponding to a spatial resolution of $\sim190\,\,\rm pc$ for 3C\,326\,N.\footnote{Convolution is performed using the Gaussian filter. At the luminosity distance to 3C\,326\,N, $1''$ corresponds to the projected distance of 1.729 kpc \citep{3c326_leftley_2024}, for which the PSF of JWST/NIRSpec amounts to $\approx 190\,\,\rm pc$.} The resulting convolved images are shown in the bottom panels. A prominent feature in all panels is the presence of a central cavity-like structure with a spatial extent of approximately 1 kpc, closely resembling the bubble morphology identified in JWST observations of 3C\,326\,N. In the convolved images, three bright clumps are also visible along the right side of the cavity, mirroring similar structures seen in the observations.\footnote{We would like to emphasize that this is a generic result and we did not aim at reaching such a close similarity.} The magenta contours represent the projected jet velocity ($\beta$) at a value of 0.7, outlining the extent and morphology of the jet within the emission maps.

\begin{figure}
	\includegraphics[scale=0.36]{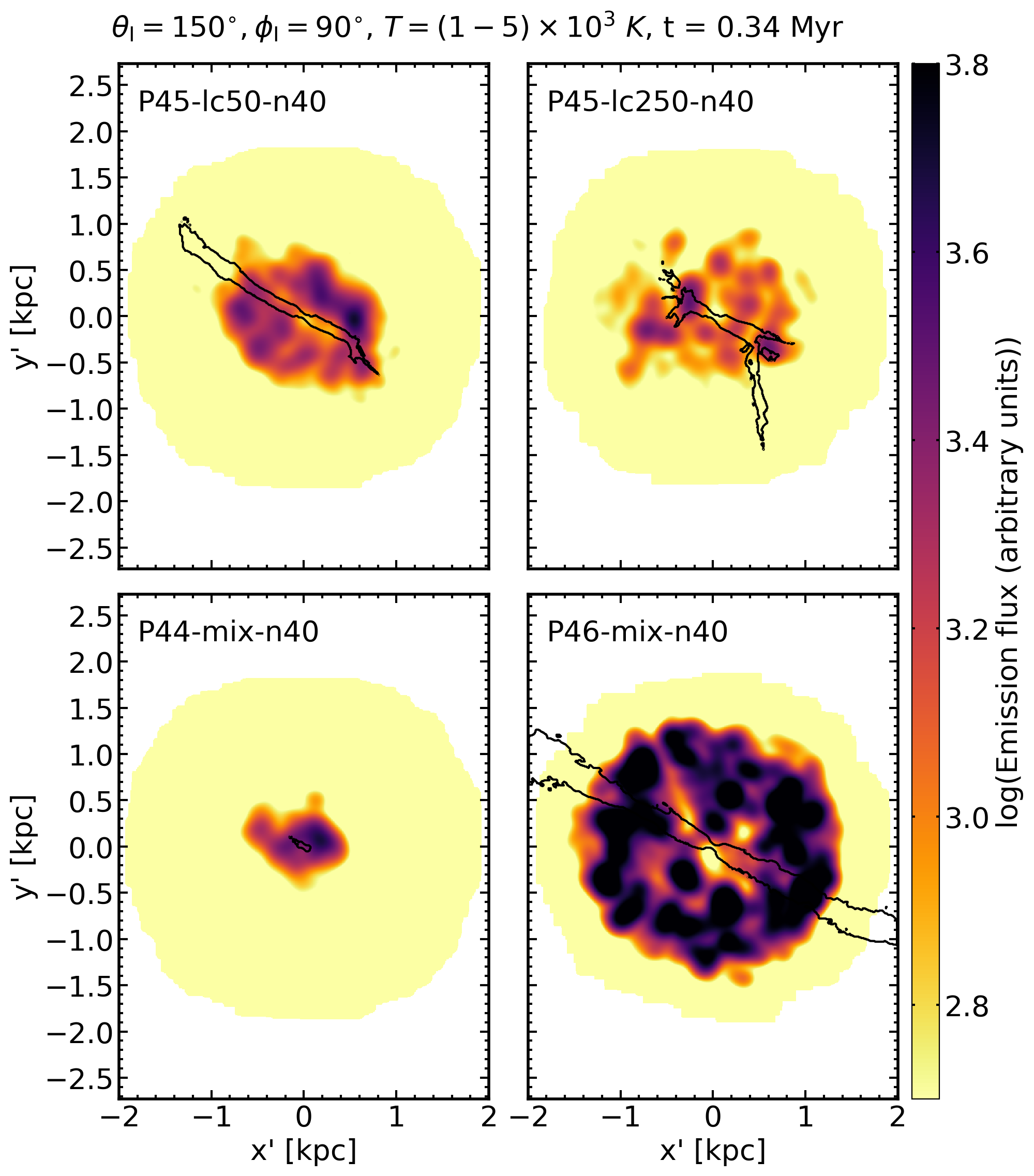}
    \caption{Synthetic emission maps, convolved with the PSF of $\rm 190\,\,pc \times190\,\,pc$, showing shocked gas in the temperature range $T = (1$--$5)\times10^{3}\,\,\rm K$ at $t = 0.34$ Myr, for the simulations P45-l100-n40 (\textit{top left}), P45-l500-n40 (\textit{top right}), P44-mix-n40 (\textit{bottom left}), and P46-mix-n40 (\textit{bottom right}), viewed along the line of sight with inclination angles $\theta_{\rm I} = 150^{\circ}$ and $\phi_{\rm I} = 90^{\circ}$. Black contours represent the projected jet $\beta$ of 0.7.}
    \label{fig:loslum_150_90_all_compare_smooth}
\end{figure}

The choice of the simulation P45-mix-n40 (with a jet power of $10^{45}\,\,\rm erg\,\,s^{-1}$ and a mixed cloud configuration) 
%combining distributions with $\lambda_{\rm max}=100\,\,\rm pc$ and $\lambda_{\rm max}=500\,\,\rm pc$, representing small and large characteristic cloud sizes) 
as the fiducial run becomes evident when compared with the convolved emission maps of other simulations listed in Table~(\ref{tab:simulation_list}), shown in Fig.~(\ref{fig:loslum_150_90_all_compare_smooth}). In the case of $l_{\rm cmax}=50\,\,\rm pc$ (P45-lc50-n40, \textit{top left}), emission appears nearly homogeneous due to the small cloud sizes and narrow inter-cloud spacing. Conversely, the $l_{\rm cmax}=250\,\,\rm pc$ model (P45-lc250-n40, \textit{top right}) produces highly clumpy and non-uniform emission, resulting from large clouds and wide separations between them. The lower-power jet (P44-mix-n40, \textit{bottom left}) fails to drive strong shocks, preventing the formation of a prominent bubble, whereas the higher-power jet (P46-mix-n40, \textit{bottom right}) disrupts the ISM too violently, quickly destroying the disk structure. Consequently, these cases do not reproduce the observed properties of 3C\,326\,N. In contrast, the P45-mix-n40 simulation successfully captures both the bubble morphology and the multiphase emission structure seen in the JWST observations, reinforcing that a mixed cloud configuration combined with an intermediate jet power provides the most realistic representation of the system.

\begin{figure*}
	\includegraphics[scale=0.35]{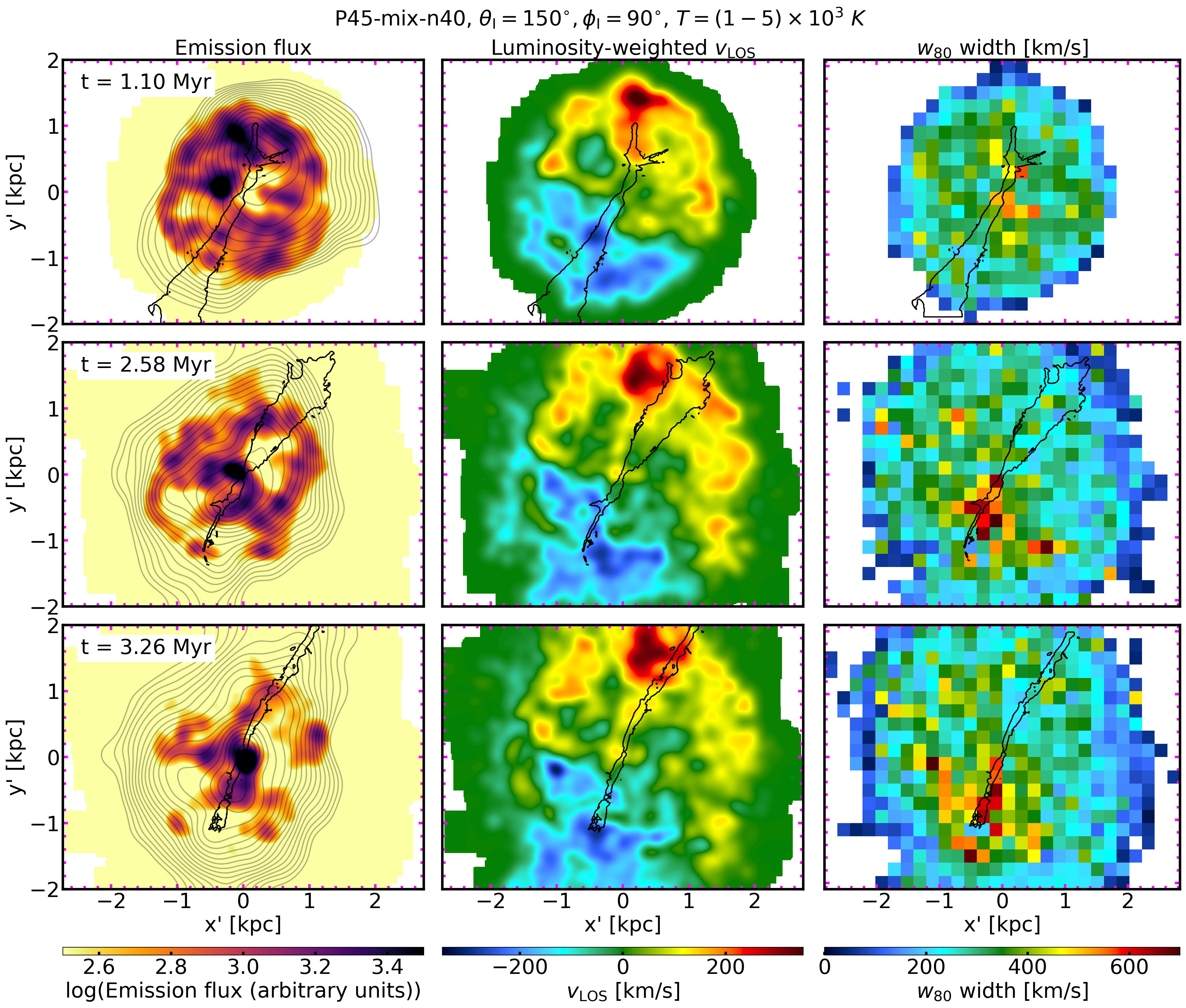}
    \caption{Convolved synthetic emission maps (\textit{left}) with overlying 9 GHz radio continuum shown as solid gray contours, luminosity-weighted LOS velocity (\textit{middle}) and $W_{80}$ widths (\textit{right}) of a shocked gas in the temperature range $T = (1$--$5)\times10^{3}\,\,\rm K$, viewed along the line of sight with inclination angles $\theta_{\rm I} = 150^{\circ}$ and $\phi_{\rm I} = 90^{\circ}$. \textit{Top to bottom:} The corresponding quantities for the epochs 1.10 Myr, 2.58 Myr, and 3.26 Myr, respectively. Black contours represent the projected jet velocity ($\beta=0.7$). The images are oriented such that the cavity and the three bright clumps lie towards the north, to match the orientation of the observational map of 3C\,326\,N (Fig.~3 of \citealt{3c326_leftley_2024}).}
    \label{fig:loslum_vlos_w80_150_90_1e3_5e3_P45_mixed_ct1t2t3_smooth}
\end{figure*}

Having established that the simulation P45-mix-n40 best explains the emission morphology of 3C\,326\,N, it is useful to study its time evolution. Fig.~(\ref{fig:loslum_vlos_w80_150_90_1e3_5e3_P45_mixed_ct1t2t3_smooth}) presents such an analysis, showing the convolved synthetic emission maps (\textit{left}), the luminosity-weighted line-of-sight (LOS) velocity (\textit{middle}), and the corresponding $W_{80}$ velocity widths (\textit{right}) at three different epochs. We also generate synthetic synchrotron surface brightness images. Since our simulations are purely hydrodynamic, 
%and we have not included particles in our simulations to track the evolution of relativistic electrons, 
we calculate the synchrotron emissivity by assuming that the magnetic energy density and energy density of relativistic electrons are fixed fractions, respectively $f_{\rm B}$ and $\eta_{\rm E}$, of the internal energy density. We adopt the nominal values of $f_{\rm B} = \eta_{\rm E} = 0.1$ in the left panels of Fig.~(\ref{fig:loslum_vlos_w80_150_90_1e3_5e3_P45_mixed_ct1t2t3_smooth}). We have shown synthetic synchrotron surface brightness contours (convolved with a Gaussian filter) at an observed frequency of 9 GHz (which falls under JVLA X-band, with a spatial resolution of $\sim 346\,\,\rm pc$), overlaying on the synthetic emission from the shocked gas. Note that there is no core component of synchrotron surface brightness, since we do not model the sub-parsec scale nuclear region of the galaxy. The images are oriented in such a way so as to match the orientation of the observational map of 3C\,326\,N as shown in the Fig.~(\ref{fig:3c326_observations}). The emission maps trace the shocked warm gas in the temperature range $T = (1$--$5)\times10^{3}\,\,\rm K$, representative of ro-vibrational $\rm H_{2}$ emission. At $t = 1.1\,\,\rm Myr$, a prominent cavity-like structure is visible, with three bright clumps forming along its northern boundary, consistent with the features seen in the JWST observations. Mass of the gas in this temperature range is $\sim 4\times10^{5}\,\,\rm M_{\odot}$, which is an order of magnitude higher than the observational estimates of warm $\rm H_{2}$ gas in 3C\,326\,N. Since the simulation was not designed to exactly replicate the system, some quantitative differences can be expected. 

The LOS velocity and $W_{80}$ maps reveal that the gas surrounding the cavity exhibits high velocity dispersions, particularly near the interface between the jet and the dense ISM, indicative of strong shear and turbulence generated by the jet–cloud interaction. The maximum LOS velocities in our simulations reach up to $\sim386\,\,\rm km\,\,s^{-1}$, which are very close to the observed maximum of $\sim380\,\,\rm km\,\,s^{-1}$ in 3C\,326\,N \citep{3c326_leftley_2024}. The $W_{80}$ map shows maximum velocity widths $\sim 560\,\,\rm km\,\,s^{-1}$ near jet-ISM interaction and typical values of $\sim 400\,\,\rm km\,\,s^{-1}$ away from the jet, consistent with the observed FWHM values $\sim 400\,\,\rm km\,\,s^{-1}$ in the disk away from the axis \citep{3c326_leftley_2024}. This suggests that the warm $\rm H_{2}$ is strongly perturbed by the jet. The remarkable similarity between various aspects such as synthetic emission morphology, ISM kinematics, etc., point towards a common physical origin of the observations and simulation results.
%The average LOS velocities in our model are closer to $\sim200\,\,\rm km\,\,s^{-1}$, making them broadly consistent with the observational constraints. 
%As the system evolves to later times ($t = 2.58$ and $3.26\,\,\rm Myr$), the cavity expands and becomes more irregular. The bright clumps become more dispersed, suggesting progressive ablation and acceleration of dense gas. 

As time progresses, the cavity becomes increasingly asymmetric, and we begin to see the emergence of smaller cavity-like features in other parts of the disk as well. These are likely the result of the jet interacting with inhomogeneous, lower-density regions within the clumpy ISM, leading to localized expansions and turbulent structures. This further emphasizes the role of ISM clumpiness in shaping the multiphase outflow morphology and sustaining jet-induced turbulence over extended timescales.

%\subsection{Simulations with high power jet ($P_{\rm jet} = 10^{46}\,\,{\rm erg\,\,s^{-1}}$)}

%For the simulation with jet power $10^{46}\,\,\rm erg\,\,s^{-1}$, we switch off the jet after $0.85\,\,\rm Myr$ of continuous activity in order to investigate the impact of intermittent jet behavior on the evolution of the ISM. This setup allows us to explore the possible outcomes of a scenario in which the jet is no longer active but its past interaction has left observable signatures in the galaxy. Since no prominent radio jet is detected in current observations of 3C\,326\,N, it is plausible that the jet was active at an earlier epoch and has since turned off, leaving behind disturbed kinematics and cavity-like structures in the multiphase ISM.

\section{Discussion and Conclusions}\label{sec.conclusions}

In this paper, we have presented a suite of 3D RHD simulations of jet--ISM interactions, exploring the parameter space of different cloud configurations (covering scales of GMCs of $\sim 50\,\,\rm pc$ to GMAs of $\sim 250\,\,\rm pc$, as well as a mixed configuration combining both scales), jet powers, and disk central densities. Our simulations incorporate a newly developed turbulence injection scheme (Appendix~\ref{appendix_A}) that provides vertical support to the disk, thereby preventing the unphysical collapse encountered in earlier studies. We have also made a direct comparison of our simulation predictions with the high-resolution JWST observations of the radio galaxy 3C\,326\,N \citep{3c326_leftley_2024}, a prototypical MOHEG where the bright H$_2$ emission is believed to be powered by jet-driven shocks.

Our key findings are summarised below:

\begin{itemize}
    \item We find that the morphology of the jet--ISM interaction and the resulting outflow properties are sensitive to the underlying cloud configuration. For a disk composed of small clouds ($l_{\rm cmax}=50$~pc), the jet propagates with little resistance provided by the ambient medium, producing widespread outflows. As a result, these simulations show the highest velocity dispersion and kinetic energy. In contrast, for large clouds ($l_{\rm cmax}=250$~pc), the jet stalls at dense clumps, leading to a more spherical bubble-like expansion and comparatively least energetic outflows among the three configurations. The mixed configuration, which provides a more realistic representation of a multi-scale ISM, captures both behaviours: one jet is stalled while the other penetrates easily. 
    This leads to outflow properties that are intermediate to the two other cases.
    %This leads to the outflows which lie intermediately between the two. 

    Analysis of the jet-driven turbulence revealed that the $l_{\rm cmax} = 250\,\,\rm pc$ case shows the highest mean compression ratio compared to the other two cases. This is because the ``flood and channel'' phase is enhanced for small-cloud and mixed configurations, which results in a lower jet power per channel. Hence the energy is predominantly dissipated through shear and mixing, resulting in solenoidal flows. In contrast, the large inter-cloud voids in the $l_{\rm cmax}=250$~pc case allow the jet to propagate with minimal fragmentation, allowing its full ram pressure to compress the massive clouds directly. 
    %These results highlight that both the amplitude and spatial structure of jet-driven outflows depend not only on the overall clumpiness of the ISM, but also on the spatial scale of its density structures.
    %To quantify the nature of jet-driven turbulence, we analysed the compression ratio $r_c = |\nabla\cdot\mathbf{v}|^{2}/(|\nabla\cdot\mathbf{v}|^{2} + |\nabla \times \mathbf{v}|^{2})$, which distinguishes solenoidal from compressible modes. In the absence of a jet, all cloud configurations exhibit nearly identical PDFs with a mean compression ratio of $\sim0.35$, indicating predominantly solenoidal turbulence. The injection of the jet, however, drives compressible outflows, with the $l_{\rm cmax}=250$~pc case showing the highest mean compression ratio of $\sim0.54$, compared to $\sim0.42$ and $\sim0.41$ for the $l_{\rm cmax}=50$~pc and mixed configurations, respectively. We attribute this difference to the contrasting interplay between the jet and the cloud distribution. In the small-cloud and mixed cases, the dense, closely spaced clouds fragment the jet into multiple channels, dissipating energy through shear and turbulent mixing, resulting in predominantly solenoidal flows. In contrast, the large inter-cloud voids in the $l_{\rm cmax}=250$~pc case allow the jet to propagate with minimal fragmentation, allowing its full ram pressure to compress the massive clouds directly.
    
    \item A comparison with \citet{wagner_2012} shows that our outflow velocities are systematically lower compared to theirs, owing to their smaller cloud sizes, spherical cloud distribution, and higher jet overpressure. Nevertheless, irrespective of density, their results converge towards ours for comparable cloud sizes (their $\rm D^{\prime\prime}_{10}$ case and our $l_{\rm cmax}=50$~pc case), suggesting that beyond a certain density threshold, feedback efficiency is determined primarily by cloud spacing rather than density.

    \item The highest power jet ($10^{46}\,\,\rm erg\,\,s^{-1}$) breaks out earliest and drives the most vigorous outflows (with mean velocities exceeding $\sim 850\,\,\rm km\,\,s^{-1}$), with velocity dispersions reaching $\sim 900\,\,\rm km\,\,s^{-1}$. It also exhibits the highest kinetic energy transfer efficiency around jet breakout, reaching up to $50\%$ of the jet power. In contrast, the lowest power jet ($10^{44}\,\,\rm erg\,\,s^{-1}$) leaves the disk largely unperturbed, with mean outflow velocities limited to $\lesssim 10\,\,\rm km\,\,s^{-1}$ and dispersions of only $\sim150\,\,\rm km\,\,s^{-1}$. We notice a general trend in all cases that once the jet is evolved out of the disk, it does not create large scale dissipation due to the effective decoupling of the jet from the host ISM. This can possibly explain the observations where in spite of the presence of a jet, the gas in the host galaxy does not show significant velocity dispersion \citep{ruffa19a,ruffa19b,ruffa20a,ruffa22a}.
    Additionally, from our accretion rate analysis, we found that even when the highest-power jet ($10^{46}\,\,\rm{erg\,\,s^{-1}}$) is switched off, the accretion flow re-forms within 2--3~Myr and reaches $\rm f_{Edd} \gtrsim 1$.

    %\item The central density of the ISM disk also plays a significant role in regulating the jet–ISM interaction. A lower mean density ($n_{\rm w0} = 10\,\,\rm{cm^{-3}}$) results in more powerful, diffuse outflows and significantly higher velocity dispersions — nearly twice those of the higher-density case ($n_{\rm w0} = 40\,\,\rm{cm^{-3}}$). This is because a lower-density ISM offers less resistance to the jet, allowing it to penetrate more easily and stir the gas more efficiently. Conversely, a denser medium absorbs more of the jet's energy locally and resists bulk acceleration. These trends are consistent with the earlier findings of \citet{wagner_2012} and \citet{mukherjee_2018}.
    
    %\item {\bf Jet confinement:}
    %The confinement of the jet depends sensitively on both cloud configuration and jet power. The mixed configuration shows the earliest breakout ($\sim 0.22$ Myr), while the large-cloud case exhibits the longest confinement ($\sim 0.28$ Myr), underscoring that the location of clouds relative to the jet path is as important as their size. Asymmetries between forward and counter jets in the mixed configuration — one encountering small clouds, the other large clouds — produce unequal arm lengths, consistent with observations of radio sources \citep{mccarthy_1993, nesvadba_2008} and earlier simulations \citep{gaibler_2011}. In terms of jet power, higher power leads to faster breakout ($<0.1$ Myr for $10^{46}\,\,\rm{erg\,\,s^{-1}}$), while lower power jets remain confined longer, depositing energy more efficiently within the disk.

    \item A key success of this work is the appearance of a jet-driven bubble in our fiducial simulation with properties akin to that observed in the radio galaxy 3C\,326\,N \citep{3c326_leftley_2024}.
    %successful reproduction of the observed jet-driven bubble in the radio galaxy 3C\,326\,N \citep{3c326_leftley_2024}. 
    Using synthetic emission maps generated in the post-processing, we find that the mixed cloud configuration with a jet power of $10^{45}\,\,\rm erg\,\,s^{-1}$ (P45-mix-n40) provides the closest match to the JWST/NIRSpec observations. The synthetic emission maps reveal a prominent central cavity of $\approx 1\,\,\rm kpc$, bounded by three bright clumps along its northern edge---features that directly mirror the observed bubble morphology in 3C\,326\,N. The LOS velocities in our fiducial simulation are very close to the observed values. The $W_{80}$ maps reveal maximum velocity widths near the jet--ISM interface, indicative of strong turbulence and shocks.
    %The maximum LOS velocities in our fiducial simulation reach $\sim 386\,\,\rm km\,\,s^{-1}$, which are very close to the observed maximum of $380\,\rm km\,\,s^{-1}$. The $W_{80}$ maps reveal maximum velocity widths $\sim 560\,\,\rm km\,\,s^{-1}$ near the jet--ISM interface, indicative of strong turbulence and shocks.
    
    In contrast, simulations with small clouds ($l_{\rm cmax}=50$~pc) produce homogeneous emission, while the large-cloud case ($l_{\rm cmax}=250$~pc) yields highly clumpy structures that do not match the observations. Similarly, lower-power jets ($10^{44}\,\,\rm erg\,\,s^{-1}$) fail to drive strong shocks, whereas higher-power jets ($10^{46}\,\,\rm erg\,\,s^{-1}$) disrupt the disk too violently, preventing the formation of a long-lived bubble. This study has allowed us to constrain the parameter space which can explain the observations of 3C\,326\,N, in which it necessitates the need to use a range of spatial scales of density structures, which is provided by our mixed cloud configuration.

    We emphasize that we did not conduct these simulations with the specific goal to match the properties of 3C\,326\,N. Instead, we used typical input values found in observed galaxies for jet power and gas properties, varying these parameters over a representative range also found in radio galaxies, and found a bubble with the right properties. This adds considerable credence to the scenario of jet-driven AGN feedback, in particular since 3C\,326\,N is a galaxy where potentially competing gas heating mechanisms like star formation and AGN radiation, that are often difficult to disentangle from the signatures of radio jet-driven AGN feedback, are very weak.
\end{itemize}

\section*{Acknowledgements}
DM, NPHN and MS acknowledge support from the IFCPAR/CEFIPRA Indo-French collaborative project ``6504-2" from 2022-2025, during which the majority of the work presented in the paper was conducted by MS, as a CEFIPRA funded Postdoctoral Fellow in IUCAA.
MS acknowledges support from the ``Zhejiang provincial top level research support program R.C.'', ``the Start-up funding of Zhejiang University'', and ``National High-Level Talent Program (Innovation Project), China''.
MM acknowledges support by the European Research Council under ERC-AdG grant PICOGAL-101019746, and support by the DFG Research Unit FOR-5195. 
MS also thanks Ankush Mandal for helpful discussions and assistance regarding the PLUTO setup. Part of the simulations and analysis were carried out using HPC facilities available at NCI Australia through various National time allocation schemes, with PI G. Bicknell, and also using HPC facilities at IUCAA.

%%%%%%%%%%%%%%%%%%%%%%%%%%%%%%%%%%%%%%%%%%%%%%%%%%
\section*{Data Availability}

No new data were generated in support of this work. The simulations
used are available from the corresponding authors upon reasonable
request.



\bibliographystyle{mnras}
\bibliography{refs} % if your bibtex file is called example.bib

%\begin{thebibliography}{99}
%\end{thebibliography}

%%%%%%%%%%%%%%%%%%%%%%%%%%%%%%%%%%%%%%%%%%%%%%%%%%

%%%%%%%%%%%%%%%%% APPENDICES %%%%%%%%%%%%%%%%%%%%%

\appendix

\section{Numerical scheme to stabilize the disk}     \label{appendix_A}

\begin{figure*}
	\includegraphics[scale=0.55]{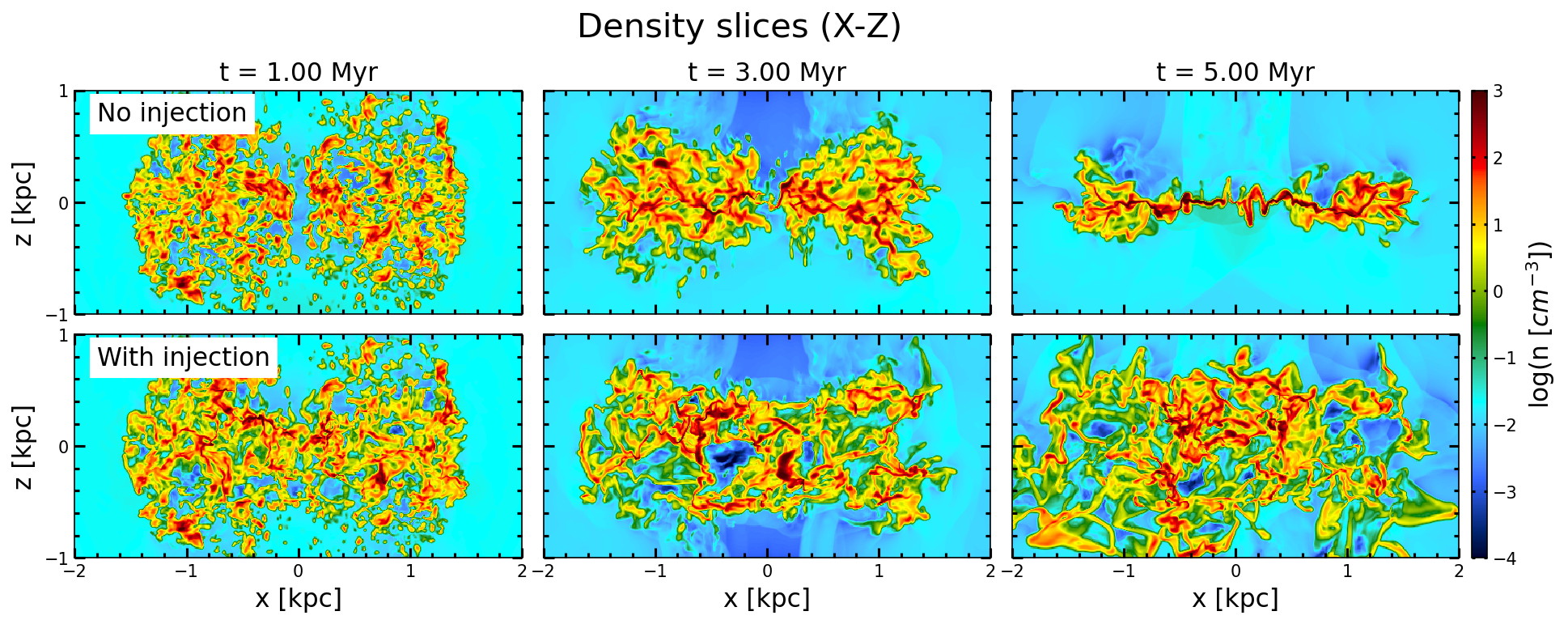}
    \caption{Density slices for no external turbulent injection  (\textit{top}) and with injection (\textit{bottom}), at 1 Myr (\textit{left}), 3 Myr (\textit{middle}), and 5 Myr (\textit{right})}
    \label{fig:rho_mixed_inj_noinj_t1t2t3}
\end{figure*}

A persistent problem in simulations following \cite{mukherjee_2018, mukherjeeIC5063_2018} has been the collapse of the dense ISM disk within 3–4 Myr due to lack of vertical support, owing to decaying turbulence in the disk. This is evident in the upper panel of Fig.~(\ref{fig:rho_mixed_inj_noinj_t1t2t3}), which shows the time evolution of an ISM disk without a jet. By 3 Myr, the disk loses significant vertical support and collapses completely by 5 Myr. The overall picture is consistent with the wide body of literature devoted to the study of ISM turbulence. \cite{mac_low_1998, stone_1998, ostriker_1999, padoan_and_nordlund_1999} study the turbulence decay rates in ISM clouds by performing HD and MHD simulations of supersonic, and super-Alfv\'{e}nic turbulence, and find that the kinetic energy decreases as a power-law in time. It is therefore essential to maintain the turbulence in the disk to  prevent its collapse.

To sustain the turbulence in the ISM, \cite{federrath_2008, federrath_2009, federrath_2010, schmidt_2009} in their high-resolution simulations, solve fluid equations with a turbulence forcing term modeled by Ornstein-Uhlenbeck (OU) process. However, such an implementation assumes an unphysical periodic boundary condition, limiting the interpretation of the simulation results to the central kiloparsec.

In our search for a method to achieve this, we initially explored a physically motivated approach by implementing stellar (and supernova) feedback using sub-grid prescriptions. However, this proved time-consuming and computationally expensive. This led us to adopt an empirically tested turbulence injection scheme, in which we artificially add velocity dispersions to the dense gas at regular time intervals. These dispersions have a Gaussian distribution and are generated using the publicly available pyFC routine.\footnote{Available at \href{https://bitbucket.org/pandante/pyfc}{https://bitbucket.org/pandante/pyfc} (created and maintained by Dr. A. Y. Wagner)} The amplitude of the injected dispersion is tuned empirically by monitoring the disk's evolution for different values and selecting the one that maintains long-term stability.

Fig.~(\ref{fig:rho_mixed_inj_noinj_t1t2t3}) illustrates the effectiveness of this approach, showing $X$–$Z$ plane density slices of the ISM disk (no jet) for two cases: without turbulent injection (upper panel) and with regularly added dispersion (lower panel). A stark difference is evident — the disk collapses without support but remains stable with turbulence injection.
%To characterize the injection, we plot in Fig.~(\ref{fig:vel_dispersion_KE_control_run_longrun}) the time evolution of the vertical velocity dispersion ($\sigma_{z}$) and kinetic energy of the dense gas, comparing both cases. 
The regular addition of dispersion at 0.1 Myr intervals helps stabilize the disk. The injection rate is $\approx 7.5\times 10^{56}\,\,\rm erg\,\,Myr^{-1}$. For comparison, 3D hydrodynamic simulations of supernova feedback by \cite{martizzi_2015} employed an injection rate of $\approx 1.4\times 10^{57}\,\,\rm erg\,\,Myr^{-1}$. This value is derived using their typical supernova-rate density of $\dot{n}_{\rm SN} = 2 \times 10^{-4}\,\,\rm SNe\,\,Myr^{-1}\,\,pc^{-3}$, scaled to our disk size. The two rates are comparable, lending confidence to the validity of our scheme.

\iffalse
\begin{figure}
	\includegraphics[scale=0.51]{images/vel_dispersion_KE_control_run_longrun.png}
    \caption{\textit{Top}: Vertical velocity dispersion ($\sigma_{z}$) of the dense ISM in control runs (no jet), comparing cases with and without turbulent injection. \textit{Bottom}: Corresponding kinetic energy evolution.}
    \label{fig:vel_dispersion_KE_control_run_longrun}
\end{figure}
\fi

\section{Volume filling factor and density PDF}

\begin{figure}
 	\includegraphics[scale=0.51]{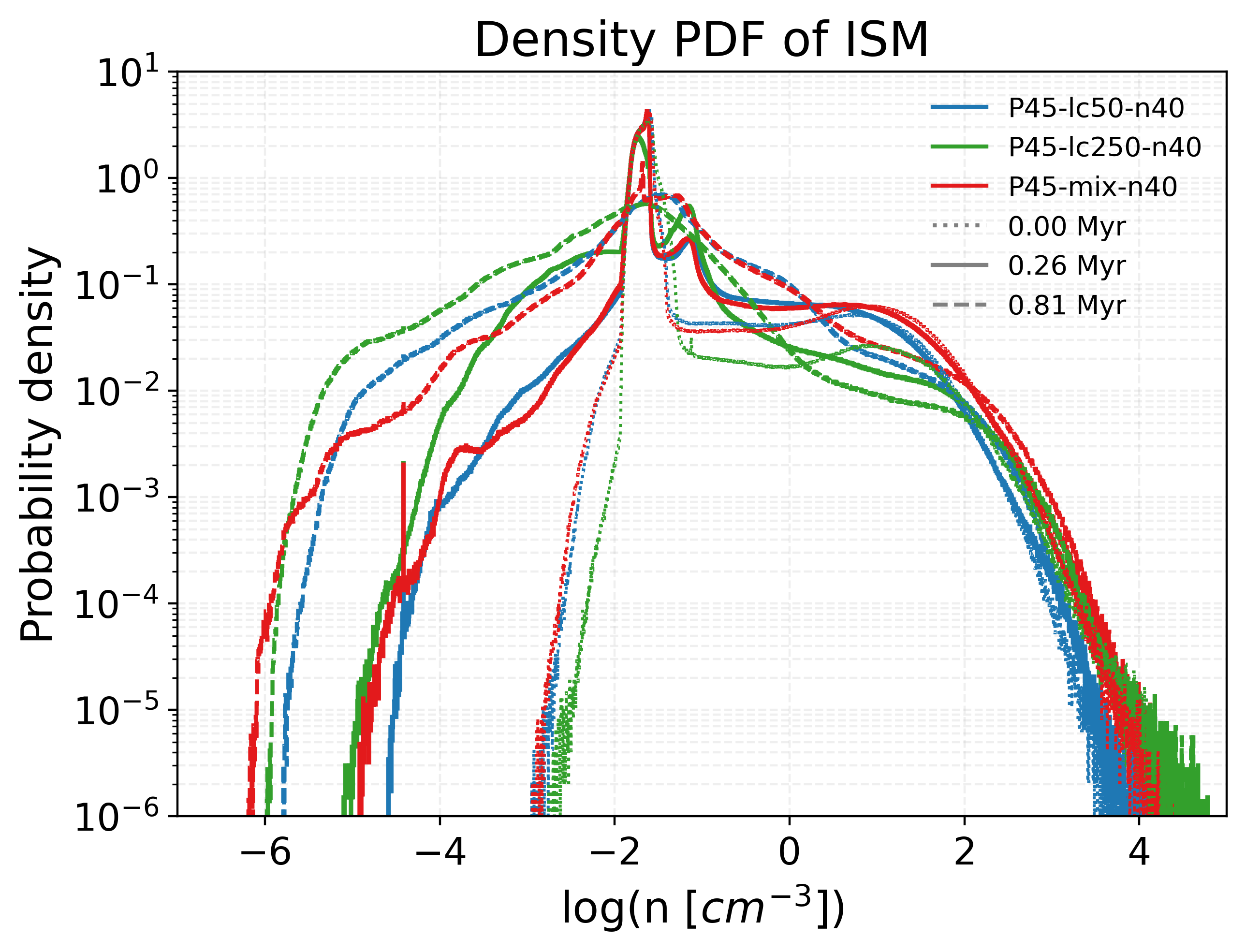}
    \includegraphics[scale=0.51]{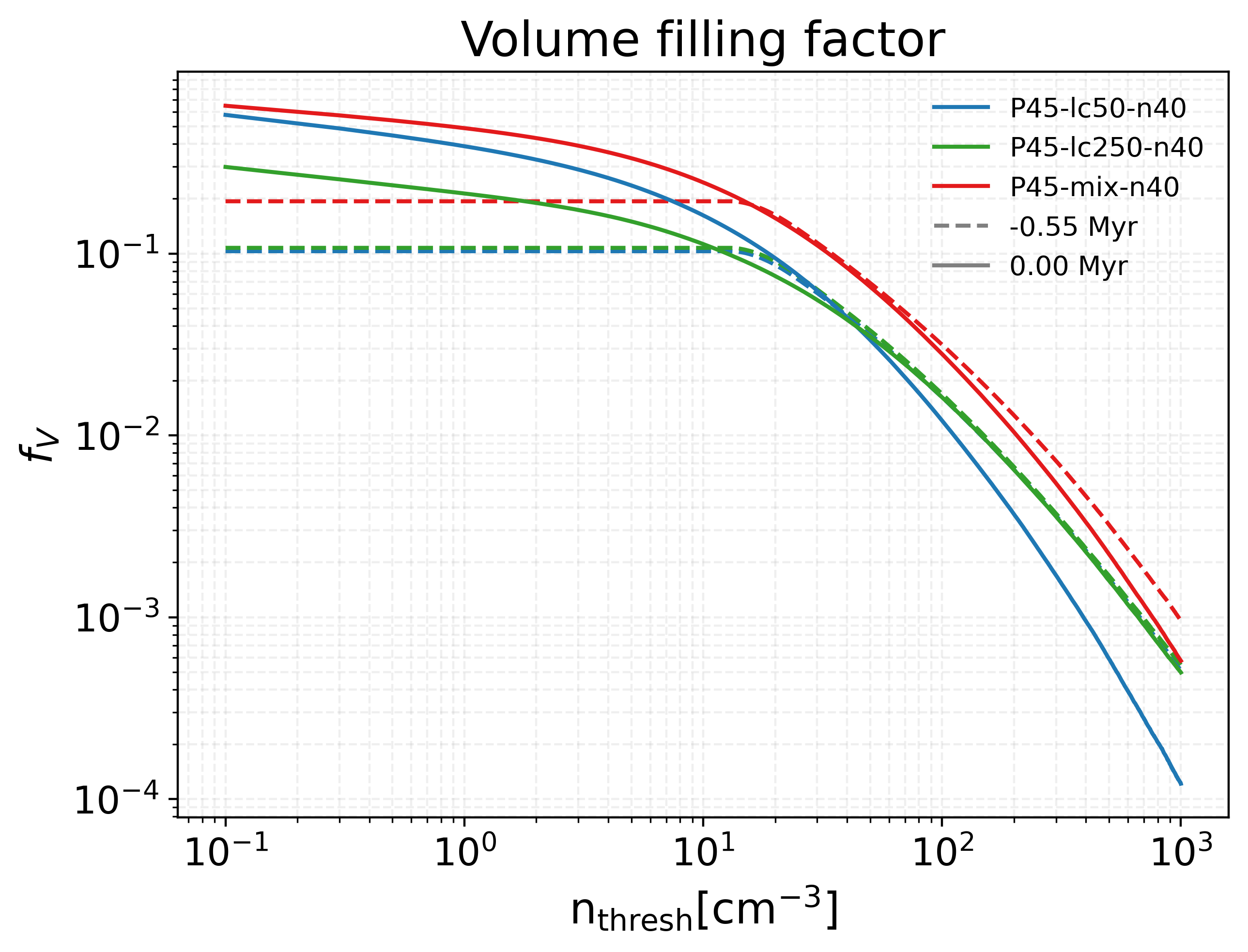}
     \caption{Top: Density PDFs at 0.26 Myr (continuous line), and 0.81 Myr (dashed line) for different cloud configurations. PDFs just before the jet injection (0 Myr) are plotted as thin dotted lines. Bottom: Volume filling factors as a function of threshold density: for the unsettled ISM (dotted lines) and settled ISM (continuous lines, just before jet injection).}
     \label{fig:rho_hist_lc_50_250_mixed}
\end{figure}

We define the volume filling factor ($f_{\rm V}$) as the fraction of total volume occupied by the gas beyond a threshold density ($n_{\rm thresh}$); the total volume being of that of the central spherical region of radius 1.5 kpc, where the warm gas is distributed. Bottom panel of Fig.~(\ref{fig:rho_hist_lc_50_250_mixed}) depicts volume filling factors of unsettled (dashed lines) and settled ISM (continuous lines), as a function of threshold density. We start off with the filling factors of 0.1 for $l_{\rm cmax} = 50, 250 \,\,\rm pc$ configurations. Since the mixed configuration has smaller clouds filled in the vacant regions between larger density structures, it has a higher filling factor ($\approx 0.2$). Before injecting the jet, we let the ISM settle for brief amount of time, the plots of which are described by continuous lines. Because of the shear, filling factor is lowered for the high density cores and consequently increases for the low density warm gas. 

Also, tracking the evolution of density probability distribution function (PDF) of the ISM helps us understand the statistics of different density structures during the jet-ISM interaction, in addition to the phase diagram presented in Sec.~(\ref{sec:phase_diagram}). Top panel of Fig.~(\ref{fig:rho_hist_lc_50_250_mixed}) shows such an evolution, with solid lines and dashed lines showing the density PDFs at 0.26 and 0.81 Myr, respectively. PDFs at the initialization are shown in thin dotted lines. We notice that the profiles remain largely unaffected for $n > 100\,\,\rm cm^{-3}$ (which constitutes dense cloud cores; see Sec.~\ref{sec:phase_diagram} for details), with a slight increase in the PDF with time. This implies that the dense cores of ISM clouds are largely unaffected. For the range $n \sim 0.1-100\,\,\rm cm^{-3}$, we see a clear separation between the PDFs, which corresponds to the varying ablation of regions around cloud cores across different cloud configurations and epochs. However, no systematic trend could be established between the PDFs for different cloud configurations in this density range. The sharp peaks around $n \sim 0.01\,\,\rm cm^{-3}$ at earlier epochs (e.g., 0.26 Myr) trace the hot halo that is not yet affected by the jet–ISM interaction. The comparatively smaller peaks around $n \sim 0.1\,\,\rm cm^{-3}$ at the same epochs can be attributed to the mixing of jet material with the outer layers of clouds. Over time, these features disappear as the jet–ISM interaction propagates through the entire computational domain. In contrast, the PDFs at the lowest densities ($n < 0.01\,\,\rm cm^{-3}$) can be ascribed to the jet material itself.

\section{Quantification of turbulence in the disk}

%mean_rc_lc50_jet = 0.422,  mean_rc_lc250_jet = 0.540,  mean_rc_mixed_jet = 0.412 
%mean_rc_lc50_nojet = 0.347,  mean_rc_lc250_jet = 0.346,  mean_rc_mixed_nojet = 0.347 
%mean_rc_P44_jet = 0.371,  mean_rc_P45_jet = 0.412,  mean_rc_P46_jet = 0.396 
%mean_rc_mixed_nojet = 0.347 

\begin{figure*}
	\includegraphics[scale=0.5]{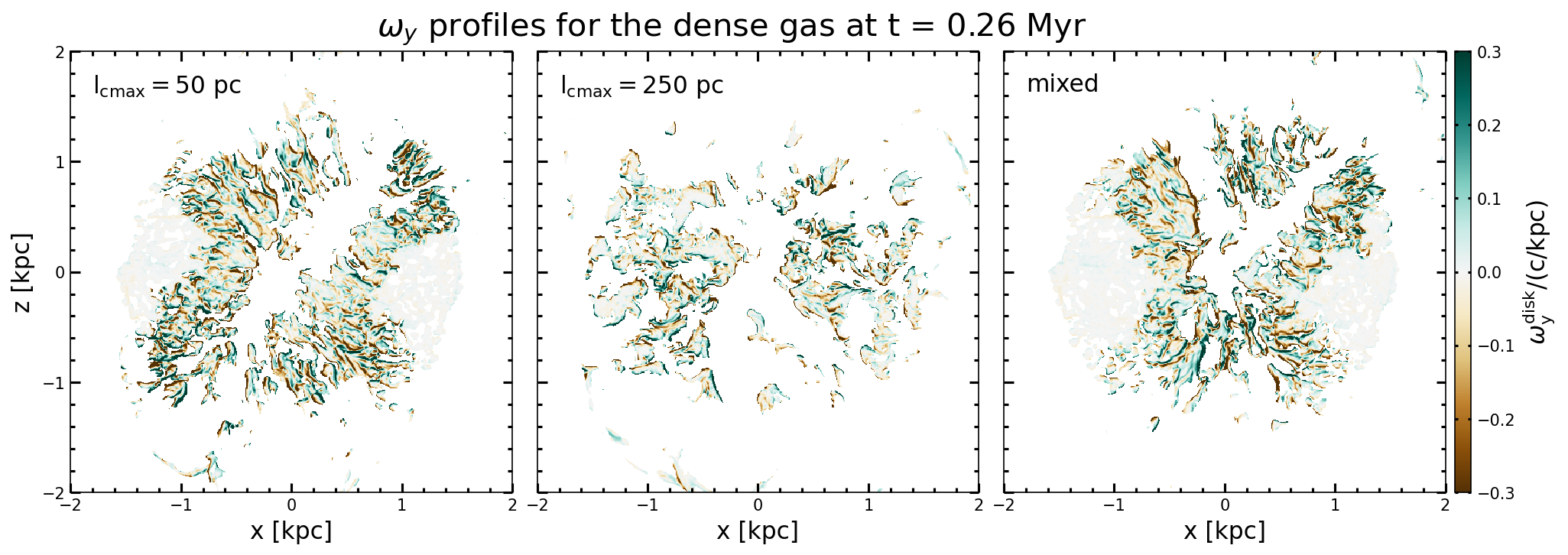}
    \includegraphics[scale=0.5]{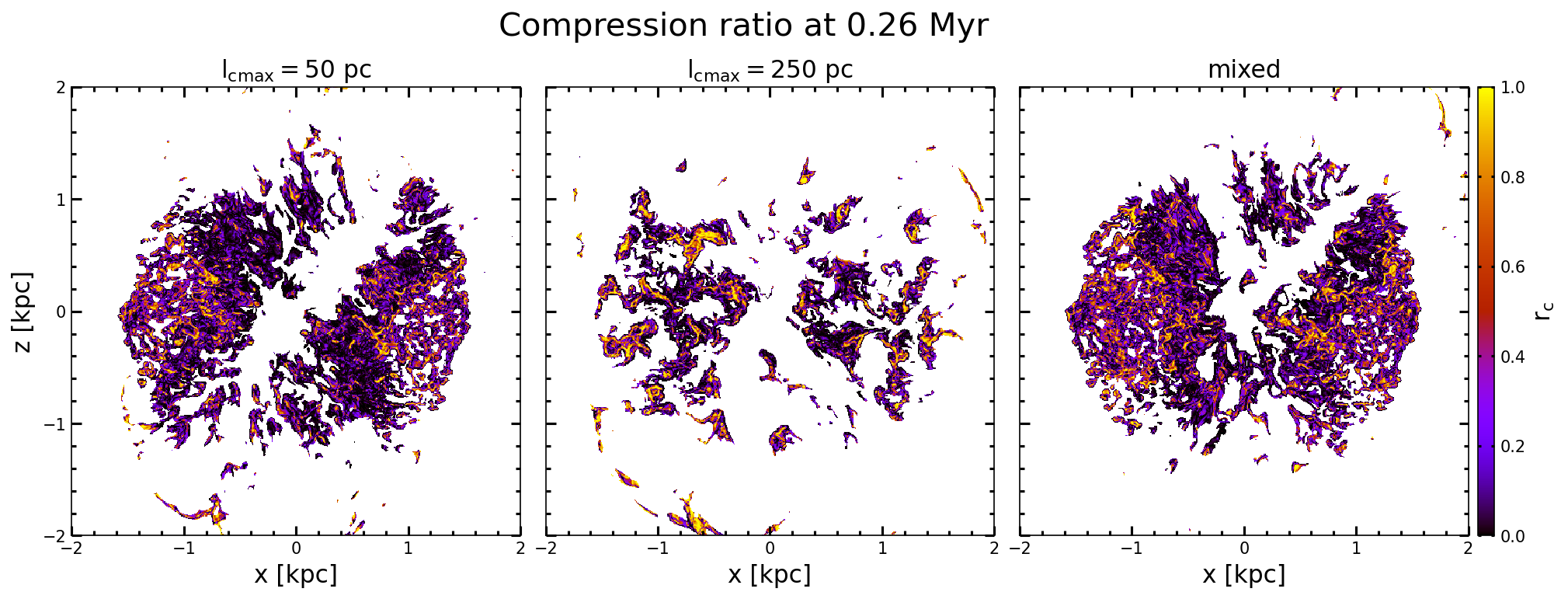}
    \caption{Top: y-component of the vorticity ($\mathbf{\omega} = \nabla \times \mathbf{v}$) corresponding to the dense gas at $t = 0.26\,\,\rm Myr$, for different cloud configurations. Bottom: Compression ratio $r_{c} = |\nabla.\mathbf{v}|^{2}/(|\nabla.\mathbf{v}|^{2} + |\nabla \times \mathbf{v}|^{2})$ for the same.}
    \label{fig:vorticity_compression_ratio_disk_t1}
\end{figure*}

Fig.~(\ref{fig:vorticity_compression_ratio_disk_t1}) presents slices of the vorticity component $\omega_y$ (upper panel) and the compression ratio $r_{c} = |\nabla\cdot\mathbf{v}|^{2}/(|\nabla\cdot\mathbf{v}|^{2} + |\nabla \times \mathbf{v}|^{2})$ (lower panel) for the dense ISM at $t = 0.26$ Myr across different cloud configurations. The details are described in Sec.(\ref{sec:compression_ratio}).

Fig.~(\ref{fig:compression_ratio_PDF_compare_jet_nojet_power}) shows the mass-weighted PDFs of the compression ratio $r_c$ for different jet powers. The lowest power case exhibits the highest PDF in the intermediate compression range ($0.1 < r_c < 0.8$), but lacks a high-$r_c$ tail, indicating that the jet is too weak to drive strong compression. The highest power case shows a bimodal distribution: a significant fraction of the gas remains in the solenoidal regime ($r_c \sim 0$), while a distinct tail extends to very high compression ratios ($r_c \sim 0.8$--$1.0$). This suggests that the jet either strongly compresses the gas or completely disrupts it into shear-dominated turbulence. The moderate power case shows the highest mean compression ratio ($\sim 0.412$), compared to $\sim0.371$ for lowest power jet and $\sim 0.396$ for the highest.%, indicating that P45 couples most efficiently with the ISM and drives the most compressible turbulence overall. This non-monotonic behaviour highlights the existence of an optimal jet power for compressible feedback, beyond which the jet energy is dissipated through cloud disruption and shear rather than compression.}

\iffalse
\begin{figure*}
	\includegraphics[scale=0.45]{images/compression_ratio_PDF_lc_50_250_mixed.png}
    \includegraphics[scale=0.45]{images/compression_ratio_PDF_P44_P45_P46.png}
    \caption{Left: PDFs of compression ratio $r_{c}$ of a dense gas (defined as $n>0.1\,\,\rm cm^{-3}$) at 0.26 Myr (thick lines), and 0.81 Myr (thin lines) for different cloud configurations. Right: Same analysis for different jet powers}
    \label{fig:compression_ratio_PDF_cloudsize_power}
\end{figure*}
\fi

\begin{figure}
	%\includegraphics[scale=0.45]{images/compression_ratio_PDF_compare_jet_nojet_cloudsize.png}
    \includegraphics[scale=0.45]{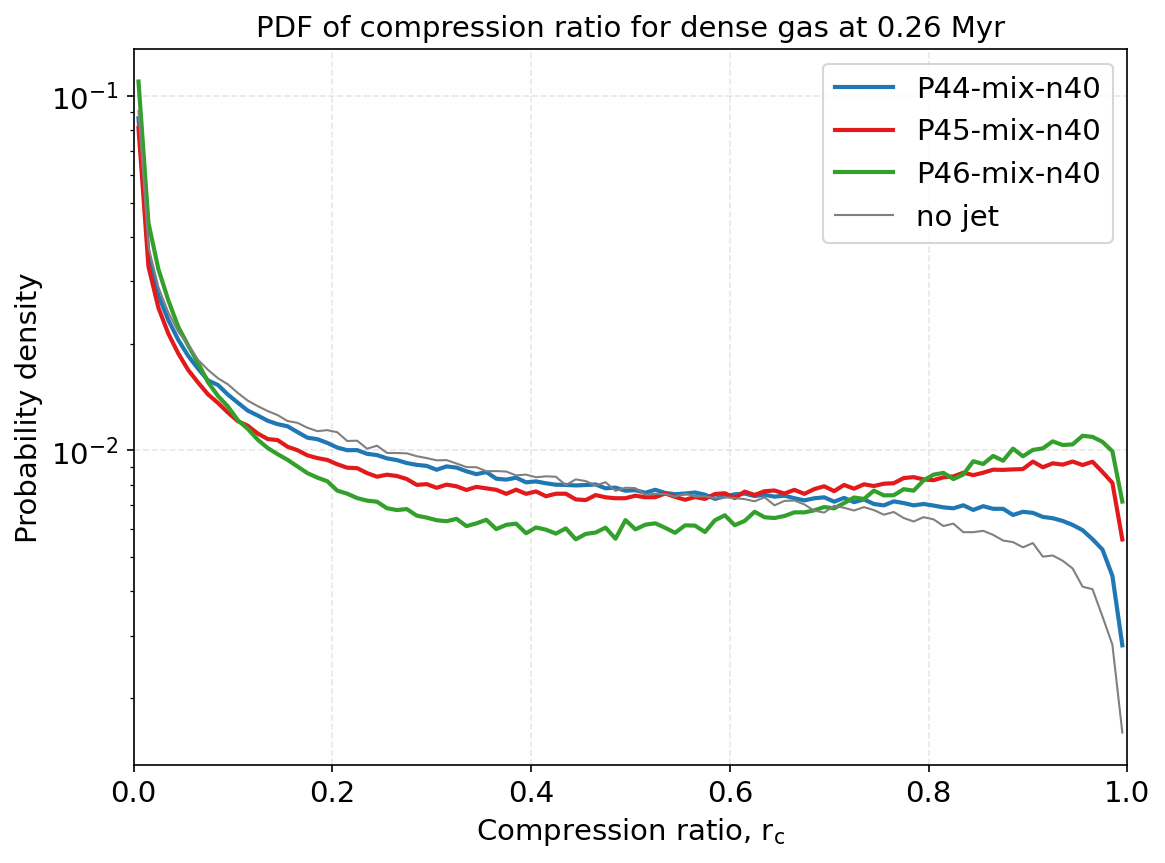}
    \caption{PDFs of compression ratio $r_{c}$ of a dense gas (defined as $n>0.1\,\,\rm cm^{-3}$) at 0.26 Myr for jetted (thick lines), and no jet (thin gray line) simulations for different jet powers.}
    \label{fig:compression_ratio_PDF_compare_jet_nojet_power}
\end{figure}

\bsp	% typesetting comment
\label{lastpage}
\end{document}